\documentclass[twocolumn,superscriptaddress,nobibnotes,footinbib,floatfix,aps,prx,citeautoscript,reprint,longbibliography]{revtex4-1}

\usepackage{graphicx}
\usepackage{epsfig}
\usepackage{subfigure}
\usepackage{color}
\usepackage{latexsym}
\usepackage{amsmath,bm,multirow}
\usepackage{amsfonts}
\usepackage{dcolumn}           
\usepackage{natbib}
\usepackage[colorlinks,linkcolor=blue,citecolor=green]{hyperref}
\usepackage{physics}
\usepackage{adjustbox}

\newcommand{\psu}{Department of Mechanical and Materials Engineering, Portland State University, Portland, OR 97201, USA}

\newcommand{\nw}{Department of Materials Science and Engineering, Northwestern University, Evanston, IL 60208, USA}
\newcommand{\nwchem}{Department of Chemistry, Northwestern University, Evanston, IL 60208, United States}
\newcommand{\argonneMS}{Materials Science Division, Argonne National Laboratory, Argonne, IL 60439, United States}

\newcommand{\yale}{Department of Applied Physics, Yale University, New Haven, CT 06511, USA}
\newcommand{\esi}{Energy Sciences Institute, Yale University, West Haven, CT 06516, USA}

\newcommand{\ustb}{School of Mathematics and Physics, University of Science and Technology Beijing, Beijing 100083, China}

\newcommand{\minikappa}{$\kappa_{\rm L}^{\rm min}$}
\newcommand{\numcompound}{2576}

\usepackage[draft]{changes} 
\usepackage[normalem]{ulem}
\definechangesauthor[name={Yi}, color=purple]{Yi}
\definechangesauthor[name={VO}, color=blue]{VO}

\usepackage{soul}

\begin{document}

\title{A unified understanding of minimum lattice thermal conductivity}

\author{Yi Xia}
\email{yxia@pdx.edu}
\affiliation{\psu}

\author{Dale Gaines II}
\affiliation{\nw}

\author{Jiangang He}
\affiliation{\ustb}

\author{Koushik Pal}
\affiliation{\nw}

\author{Mercouri G. Kanatzidis}
\affiliation{\nwchem, \argonneMS}

\author{Vidvuds Ozoli\c{n}\v{s}}
\email{vidvuds.ozolins@yale.edu}
\affiliation{\yale, \esi}

\author{Chris Wolverton}
\email{c-wolverton@northwestern.edu}
\affiliation{\nw}

\date{\today}

\begin{abstract}

We propose a first-principles model of minimum lattice thermal conductivity ($\kappa_{\rm L}^{\rm min}$) based on a unified theoretical treatment of thermal transport in crystals and glasses. We apply this model to thousands of inorganic compounds and discover a universal behavior of $\kappa_{\rm L}^{\rm min}$ in crystals in the high-temperature limit: the isotropically averaged $\kappa_{\rm L}^{\rm min}$ is independent of structural complexity and bounded within a range from $\sim$0.1 to $\sim$2.6 W/[m$\cdot$K], in striking contrast to the conventional phonon gas model which predicts no lower bound. We unveil the underlying physics by showing that for a given parent compound \minikappa\ is bounded from below by a value that is approximately insensitive to disorder, but the relative importance of different heat transport channels (phonon gas versus diffuson) depends strongly on the degree of disorder. Moreover, we propose that the diffuson-dominated \minikappa\ in complex and disordered compounds might be effectively approximated by the phonon gas model for an ordered compound by averaging out disorder and applying phonon unfolding. With these insights, we further bridge the knowledge gap between our model and the well-known Cahill-Watson-Pohl (CWP) model, rationalizing the successes and limitations of the CWP model in the absence of heat transfer mediated by diffusons.  Finally, we construct graph network and random forest machine learning models to extend our predictions to all compounds within the Inorganic Crystal Structure Database (ICSD), which were validated against thermoelectric materials possessing experimentally measured ultralow $\kappa_{\rm L}$. Our work offers a unified understanding of  $\kappa_{\rm L}^{\rm min}$, which can guide the rational engineering of materials to achieve $\kappa_{\rm L}^{\rm min}$.

\end{abstract}

\maketitle

\section{Introduction}

 Knowing the lower limit to the lattice thermal conductivity of crystals is of fundamental interest and technological importance, particularly relevant to thermal energy conversion and management applications~\cite{Bell1457}. The minimum lattice thermal conductivity (\minikappa), a concept first proposed by Slack~\cite{slackbook2}, sets the upper limit of thermoelectric conversion efficiency for a given electronic transport profile~\cite{slackbook, Snyder2008,Sootsman2009,HeJian2017} and provides the maximum thermal insulation for a substrate used in thermal barrier coatings~\cite{CLARKE200367,Thakare:2020aa}. The fundamental physics of heat transport in crystals at the lower limit of thermal conductivity continues to be an active research topic~\cite{Cahill1992,slackbook2,CLARKE200367,Agne2018}.

Motivated by the observation that the experimentally accessible \minikappa\ is often found in amorphous solids, the initial attempts to understand \minikappa~\cite{Cahill1988,Cahill1989} relied on a model proposed by Einstein~\cite{Einstein1911}, who assumed that the mechanism of heat transport in crystals was a random walk of the thermal energy between neighboring atoms vibrating with random phases. Einstein's random walk model was later generalized by Cahill, Watson, and Pohl~\cite{Cahill1992} (referred to as the CWP model) to incorporate the Debye model of vibrations by adopting a wavelength-dependent mean free path (MFP). An earlier approach by Slack \cite{slackbook2} as well as the more recent study by Clarke \cite{CLARKE200367} also assumed that phonons are the dominant heat carriers, but used different assumptions about the phonon MFP. The CWP, Slack, and Clarke theories of \minikappa\  are all based on the conventional phonon gas model (PGM) and the Boltzmann gas kinetic equation~\cite{kittel2004introduction}, where the primary heat carriers are propagating phonons~\cite{Agne2018}. On the other hand, recognizing the potential failure of the PGM in the disordered regime, Allen and Feldman~\cite{AllenPRL1989,Allen1993disorder} developed the diffuson theory of heat transport in disordered solids. The Allen-Feldman theory is based on Kubo-Greenwood formula in the harmonic approximation and the primary heat carriers are diffusons, which are described by the off-diagonal terms of the heat flux operator. Recently, a phenomenological \minikappa\ model was developed by Agne, Hanus, and Snyder~\cite{Agne2018} based on the diffuson theory, aiming to resolve the overestimation of \minikappa\ from the CWP model in comparison to experiments.

Despite significant advances, a comprehensive understanding of \minikappa\ is still elusive. A crucial yet missing piece of the puzzle is a first-principles-based theory that accurately describes \minikappa\ over the whole spectrum of materials between the prototypical classes of simple crystals and disordered amorphous solids. The development of such an advanced model would shed light on crucial open questions. For instance, the heat transfer pathways mediated by localized diffusons and propagating phonons are often classified into independent channels~\cite{allen1999diffusons,Mukhopadhyay1455,Luo:2020aa,PhysRevLett.125.085901,Hanus2021}. Is there a unified physical picture of the two different kinds of heat carriers when approaching \minikappa? Why does the CWP model work remarkably well for solids with strong disorder, despite retaining the PGM nature and neglecting all optical phonons~\cite{Cahill1988,Cahill1989,Cahill1992}? There are recent experimental discoveries of crystalline compounds with ultralow and glass-like lattice thermal conductivity $\kappa_{\rm L}$ that resembles $\kappa_{\rm L}^{\rm min}$~\cite{AgTlTe2005,Liu:2012aa,Lu2013,CsAg5Te32016,LIN2017816}. What is the general behavior of $\kappa_{\rm L}^{\rm min}$ in crystals, i.e., how does it vary with structural complexity and atomic composition? Ultimately, is there a physical lower or upper bound to $\kappa_{\rm L}^{\rm min}$?


\section{Results and discussions}

\begin{figure*}[htp]
	\includegraphics[width = 1.0\linewidth]{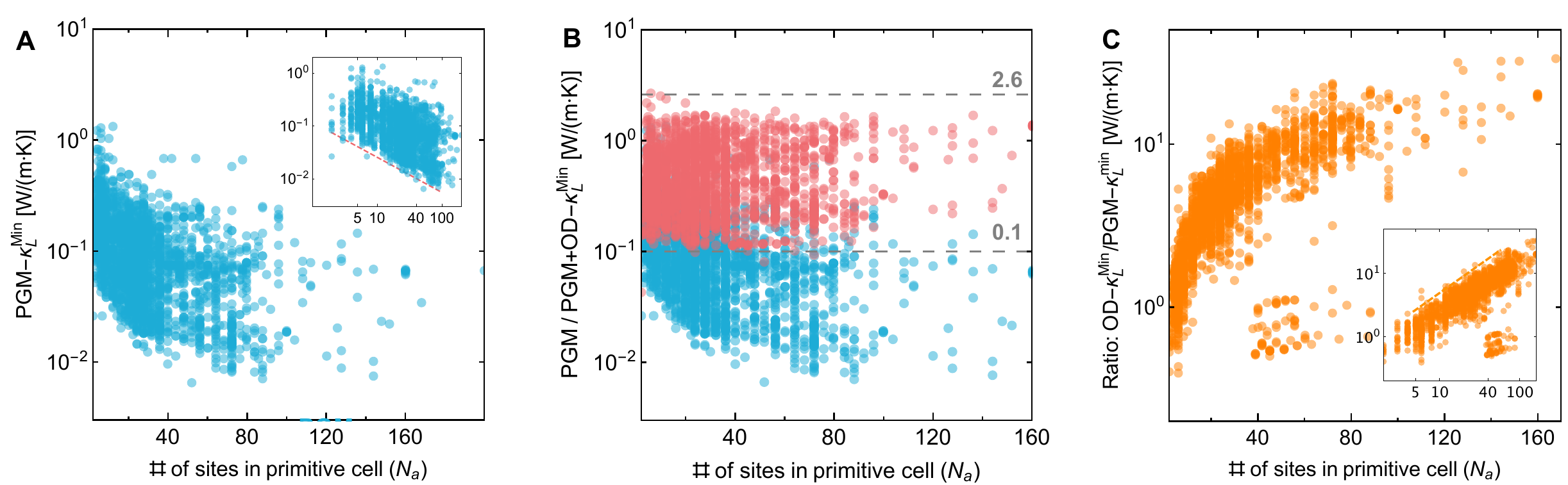}
	\caption{ 
	Contributions to minimum lattice thermal conductivity from the phonon gas model and the off-diagonal terms based on Eq.[\ref{eq:minikappa}]. (A) Calculated minimum lattice thermal conductivity (\minikappa) as a function of the number of atomic sites ($N_{\rm a}$) in a primitive cell using the phonon gas model (denoted as PGM-\minikappa), i.e., the diagonal part of Eq.(\ref{eq:minikappa}). The inset is plotted with logarithmic scales and the dashed line is plotted following the scaling of $N_{\rm a}^{-2/3}$ obtained from Slack's model~\cite{slack1973nonmetallic}, showing the monotonic decay of PGM-\minikappa\ with increasing $N_{\rm a}$. (B) The same as (A) but with additional red disks showing the calculated \minikappa\ accounting for both the PGM and off-diagonal (OD) contributions, denoted as PGM+OD-\minikappa. The gray dashed lines indicate values of 0.1 and 2.6 W/[m$\cdot$K] for the lower and the upper bounds of PGM+OD-\minikappa, respectively. (C) The ratio of OD-\minikappa\ and PGM-\minikappa\ as  a function of $N_{\rm a}$. The inset is displayed with logarithmic scales, showing the increasing trend of OD-\minikappa/PGM-\minikappa. All results were obtained at 600~K.
	}
	\label{fig:minikappa}
\end{figure*}

Recent theoretical advances towards a unified theory of thermal transport in crystals and glasses have enabled a consistent treatment of different heat carriers such as propagating phonons and localized diffusons \cite{Simoncelli2019,Isaeva2019}. Herein, we construct a \minikappa\ model based on the unified theory of Simoncelli, Marzari, and Mauri~\cite{Simoncelli2019,wigner2022} by adopting the approximation that  each phonon mode's lifetime is equal to one half of its vibrational period, $\tau_\mathbf{q}^{s}=\pi/\omega_\mathbf{q}^{s}$. This assumption follows the original proposal by Einstein~\cite{Einstein1911} which was later adopted by the CWP model~\cite{Cahill1992}. We choose not to use the minimum MFP criteria (e.g., by assuming that MFP is equal to the smallest atomic spacing) because for optical phonon modes with vanishing group velocities it will lead to infinite phonon lifetimes, which is unphysical and tends to diminish their off-diagonal contribution to thermal conductivity (see below). Using the formula for thermal conductivity from Ref.~\cite{Simoncelli2019}, we arrive at the following expression:
\begin{equation}
\begin{split}
\kappa_{\rm L}^{\rm min}  = & \frac{\pi\hbar^2}{k_{\rm B}T^2VN_{\mathbf{q}}} \sum_{\mathbf{q}}\sum_{s, s^{\prime}} \frac{ (\omega_{ \mathbf{q}}^{s}+\omega_{\mathbf{q}}^{s^{\prime}} )^2 }{2} \mathbf{v}_\mathbf{q}^{s,s^{\prime}} \otimes \mathbf{v}_\mathbf{q}^{s^{\prime},s}  \\
& \cdot \frac{ \omega_\mathbf{q}^{s}n_\mathbf{q}^{s} ( n_\mathbf{q}^{s}+1) + \omega_\mathbf{q}^{s^{\prime}}n_\mathbf{q}^{s^{\prime}} ( n_\mathbf{q}^{s^{\prime}}+1) }{ 4\pi^2(\omega_\mathbf{q}^{s^{\prime}}-\omega_\mathbf{q}^{s})^2 + (\omega_\mathbf{q}^{s}+\omega_\mathbf{q}^{s^{\prime}})^2 }
\end{split}
\label{eq:minikappa}
\end{equation}
where $\hbar$, $k_{B}$, $T$, $V$, $N_{\mathbf{q}}$ are, respectively, the reduced Planck constant, the Boltzmann constant, the absolute temperature, the volume of the unit cell, and the total number of sampled phonon wave vectors. Phonon modes are denoted by the wave vector $\mathbf{q}$ and mode index $s$. The key quantities entering Eq.[\ref{eq:minikappa}] are phonon mode-resolved frequencies $\omega_\mathbf{q}^{s}$, population numbers $n_\mathbf{q}^{s}$, and the generalized group velocity tensors $\mathbf{v}_\mathbf{q}^{s,s^{\prime}}$, with the latter calculated as \cite{Allen1993disorder,Simoncelli2019}
\begin{equation}
\begin{split}
    \mathbf{v}_\mathbf{q}^{s,s^{\prime}} = & \frac{i}{\omega_\mathbf{q}^{s}+\omega_\mathbf{q}^{s^{\prime}}} \sum_{\alpha,\beta}\sum_{m,p,q}e_\mathbf{q}^{s}(\alpha,p) D^{pq}_{\beta\alpha}(0,m) \\
    & \times (\mathbf{R}_m+\mathbf{R}_{pq}) e^{i\mathbf{q}\cdot\mathbf{R}_m} e_\mathbf{q}^{s^{\prime}}(\beta,q).
\end{split}
\label{eq:groupv}
\end{equation}
Here, $e$, $D$, and $\mathbf{R}$ denote the polarization vector, the dynamical matrix, and the lattice vector, respectively. $\alpha$/$\beta$, $m$, and $p$/$q$ are indices labeling the Cartesian coordinate, the unit cell, and the atoms within the unit cell. 

Physically, we can further decompose \minikappa\ into two parts: the diagonal part with $s=s^{\prime}$, which corresponds to the PGM, denoted as PGM-\minikappa, and the off-diagonal part, which accounts for the diffuson channel, denoted as OD-\minikappa. The total \minikappa\ will be referred to as PGM+OD-\minikappa. Immediately, we see that the evaluation of Eq.[\ref{eq:minikappa}] only requires the knowledge of the harmonic phonon dispersion, making it amenable to large-scale DFT calculations. We applied Eq.[\ref{eq:minikappa}] to a selected set of \numcompound~inorganic compounds, taking advantage of the tabulated harmonic force constants within the phonon database (phononDB) generated by Togo {\it et al.\/}~\cite{phonondbtogo,Togo2015} using crystalline structures from the Materials Project (MP)~\cite{matproj,pymatgen,mpapi}. 


Temperature enters Eq.~(\ref{eq:minikappa}) both explicitly in the prefactor and implicitly via the phonon occupation numbers $n_\mathbf{q}^{s}$. Because \minikappa\ increases monotonically with increasing temperature up to the Debye temperature as a result of enhanced heat capacity~\cite{Cahill1992}, and \minikappa\ in crystals is usually approached at high temperatures, we focus on exploring the high-temperature regime. We performed calculations at $T=300$, 600, and 900~K. We find that \minikappa\ of most compounds calculated at 600~K exhibit considerable increment over those calculated at 300~K, whereas \minikappa\ varies by less than 5\% between $T=600$ and 900~K for approximately 95\% of the studied compounds, with the largest decrease being only 14\%. Considering the larger availability of experimental measurements of $\kappa_{\rm L}$ at 600~K than 900~K, we will focus on the $T$ = 600~K results in the following discussion. For anisotropic crystals, the components of \minikappa\ are further averaged along the three principle crystallographic axes. We refer the readers to the Methods section for a more detailed discussion of the structure types, chemical compositions, and criteria for numerically converging \minikappa.

The calculated values of PGM-\minikappa, OD-\minikappa, and PGM+OD-\minikappa\ as functions of the number of atomic sites in the primitive cell, $N_{\rm a}$, are plotted in Fig.\ref{fig:minikappa}. We associate the value of $N_{\rm a}$ with the degree of structural complexity of a compound. We see from Fig.\ref{fig:minikappa}A that PGM-\minikappa\ covers a large range of values, spanning from approximately 0.01 to 1 W/[m$\cdot$K] in the high temperature limit (600~K and above). There is a clear trend that PGM-\minikappa\ decreases monotonically with increasing $N_{\rm a}$, which can be clearly seen from the log-log plot in the inset of Fig.\ref{fig:minikappa}A). In contrast, the total PGM+OD-\minikappa\ in Fig.\ref{fig:minikappa}B exhibits a much narrower spread from approximately 0.1 to 2.6~W/[m$\cdot$K], independent of $N_{\rm a}$. The lower (0.1~W/[m$\cdot$K]) and the upper (2.6 ~W/[m$\cdot$K]) bounds were determined by analyzing the frequency distribution histogram of all calculated \minikappa, as detailed in the Supplementary Materials section. There are only four compounds within our calculations having values of PGM+OD-\minikappa\ outside these bounds. The largest deviation from the lower bound is found in crystalline $P6(3)/mmc$-Ar, which has a value of 0.043 W/[m$\cdot$K], and all the other compounds have values larger than 0.081~W/[m$\cdot$K]. Meanwhile, for the upper bound, the largest PGM+OD-\minikappa\ is associated with $P4(2)/mnm$-TiO$_{2}$, with a value of 2.64~W/[m$\cdot$K]. The comparison of PGM-\minikappa\ and PGM+OD-\minikappa\ reveals the increasing importance of the off-diagonal diffuson term OD-\minikappa\ with increasing structural complexity $N_{\rm a}$. This is further illustrated by the ratio of OD-\minikappa\ to PGM-\minikappa\ in Fig.\ref{fig:minikappa}C. We see that OD-\minikappa\ quickly surpasses PGM-\minikappa\ above $N_{\rm a} \sim 10$ and OD-\minikappa\ becomes an order of magnitude higher than PGM-\minikappa\ above $N_{\rm a} \sim 60$. We also observe a group of compounds (e.g., $P3m1$-ZnS) with ratios of OD-\minikappa\ to PGM-\minikappa\ significantly smaller than the majority trendline. Our analysis of their structures reveals that these compounds display extreme crystalline anisotropy, i.e., the lattice constant of the primitive cell along one axis is exceedingly large and thus $N_{\rm a}$ is large. However, for these compounds, the ratios of OD-\minikappa\ to PGM-\minikappa\ are reduced along the other two axes with much smaller lattice constants, thus giving rise to overall reduced ratios after averaging over the three spatial directions.

The monotonic decay of PGM-\minikappa\ as a function of $N_{\rm a}$ is not unexpected. As already reflected in the Slack model~\cite{slack1973nonmetallic,slack1979thermal}, $\kappa_{\rm L}$ based on the PGM is proportional to $N_{\rm a}^{-2/3}$ provided that acoustic phonon modes with high group velocities dominate $\kappa_{\rm L}$. The monotonic decay of PGM-\minikappa\ can be attributed to the fact that the increase in $N_{\rm a}$ leads to a reduced Brillouin zone that effectively folds back the high-energy acoustic modes into the first Brillouin zone as optical modes, resulting in both suppressed spectral weight and reduced group velocity. It is worth noting that strategies based on the above argument have been successfully used to search for complex materials with intrinsically low lattice thermal conductivity~\cite{toberer,LIN2017816}. However, our results show that these strategies may be no longer effective when phonons are strongly scattered by anharmonicity or disorder when the heat transport is dominated by the OD terms beyond the PGM. This is especially true for systems with $\kappa_{\rm L}$ approaching \minikappa. In this scenario, the breakdown of the PGM is due to the crossover in the heat transport from propagating phonons to localized diffusons, as initially proposed by Allen and Feldman~\cite{AllenPRL1989,Allen1993disorder}. As a result, the total PGM+OD-\minikappa\ provides a much more reasonable estimation of \minikappa\ compared to the phonon term PGM-\minikappa\ alone. Importantly, despite such a crossover from PGM-\minikappa\ to OD-\minikappa, the bounds of the total PGM+OD-\minikappa\ remain independent of $N_{\rm a}$. This suggests an effective interconversion between the two fundamentally different heat transfer channels, which might be induced by variations in chemical composition, atomic disorder, lattice distortion, structural complexity ($N_{\rm a}$), and other factors. This invites the following two questions: (i) what are the key factors that determine the interconversion between PGM-\minikappa\ and OD-\minikappa\ across various compounds? and (ii) is it possible to revert such a interconversion, for example, via approximating PGM+OD-\minikappa\ of compounds with large $N_{\rm a}$ using only PGM-\minikappa?

\begin{figure*}[htp]
	\centering
	\includegraphics[width = 1.0\linewidth]{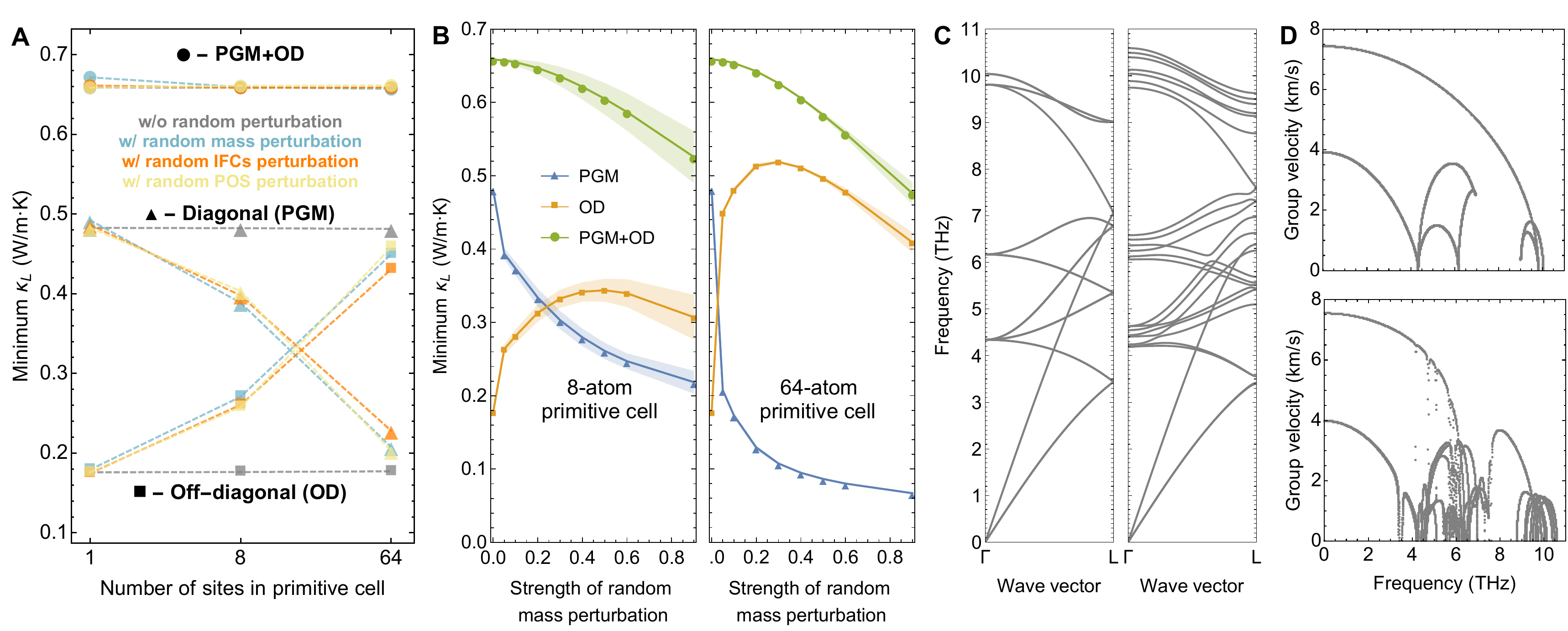}
	\caption{
    Impacts of disorder and structural complexity on minimum lattice thermal conductivity. (A) Calculated total \minikappa\ (PGM+OD-\minikappa) and the corresponding decomposed contributions (PGM-\minikappa\ and OD-\minikappa) of a model system (fcc-Al) as functions of the number of atomic sites ($N_{\rm a}$) in the designed primitive cell and the presence of various kinds of disorder. We artificially create three kinds of disorder in the perfect crystal of fcc-Al by applying small random perturbations (within 5\% of their original values) to atomic mass, interatomic force constants (IFCs), and atomic position (POS), denoted in blue, orange, and yellow, respectively (with dashed line as a guide to the eye). (B) Calculated PGM+OD-\minikappa(green disks), PGM-\minikappa(blue triangles), and OD-\minikappa(orange squares) of fcc-Al as a function of the strength of mass disorder, realized by applying random percentagewise mass perturbation. For example, 20\% random mass perturbation means a random mass ranging from -20\% and 20\% of the original atomic mass of the element is added to each atom. The left and right panel show the results obtained for an 8-atom and 64-atom primitive cell, respectively. The shaded areas indicate the uncertainty obtained by averaging many independent random mass perturbations. (C) Phonon dispersions of the perfect fcc-Al (left panel) and an 8-atom primitive cell with 20\% random mass perturbations (right panel). (D) Phonon group velocities of the perfect fcc-Al (upper panel) and an 8-atom primitive cell with 20\% random mass perturbations (lower panel). The \minikappa\ shown in panel A and B were all calculated at 300~K.
	}
	\label{fig:disorder}
\end{figure*}

Answering the above two questions will help establish a unified understanding of \minikappa. The challenge arises from the fact that the crossover in the heat transport from phonons to diffusons is observed in a large set of compounds with diverse characteristics. To reveal and disentangle the impacts of chemical composition, interatomic interaction, and structural complexity on \minikappa, we use a simple elemental structure, i.e., face-centered cubic aluminium (fcc-Al), as a base model and then manually generate perturbations on this model to mimic the variations in structures and chemistries among different compounds. Specifically, random perturbations are applied on the atomic mass, interatomic force constants (IFCs), and atomic position. We also vary the number of atoms ($N_{\rm a}$) in the unit cell by repeating the primitive cell of fcc-Al. As a consequence, we can use these models to approximate a variety of situations ranging from ordered to disordered by varying the strength of perturbations and $N_{\rm a}$. Note that we strictly enforce physical constraints when applying these perturbations, such as the translational invariance of the crystal.

Fig.\ref{fig:disorder}A summarizes the calculated total \minikappa\ and its decompositions (PGM and OD) when weak perturbations are applied respectively to atomic masses, IFCs, and atomic positions in unit cells with increasing $N_{\rm a}$. First, we observe that neither the total \minikappa\ nor its decomposition change by simply increasing $N_{\rm a}$ without introducing disorder. This is as expected because nothing has been changed physically and merely a larger unit cell is used as the primitive cell for the perfect fcc-Al crystal. In contrast, when small perturbations are applied, we notice different trends in PGM-\minikappa\ and OD-\minikappa\ across the unit cells with increased $N_{\rm a}$: (i) when $N_{\rm a}$=1, perturbations lead to only small changes in both PGM-\minikappa\ and OD-\minikappa, and the PGM-\minikappa\ dominates over the OD-\minikappa, as expected for a simple crystal; (ii) when $N_{\rm a}$=8, perturbations start to convert PGM-\minikappa\ to OD-\minikappa, with the former still larger than the latter; (iii) when $N_{\rm a}$=64, the values of PGM-\minikappa\ and OD-\minikappa\ are nearly exchanged, and OD-\minikappa\ becomes the major contribution to the total \minikappa. Surprisingly, the total \minikappa\ largely remains constant with both varying perturbations and $N_{\rm a}$.

The above results reveal that the total \minikappa\ is not sensitive to small perturbations of atomic masses, positions, and IFCs of a given base crystal structure. In the other words, the capability of lattice heat transfer of a given system that approaches maximum phonon scattering ($\tau_\mathbf{q}^{s} \approx \pi/\omega_\mathbf{q}^{s}$) can neither be reduced nor enhanced significantly with small adjustments in atomic compositions, crystal structure, and interatomic interactions. This is in contrast to the case of weak phonon scattering in some materials (e.g. diamond and BN), wherein small mass perturbations such as isotope scattering can lead to a considerable reduction in $\kappa_{\rm L}$~\cite{kechenBN}. Conversely, the relative contributions of PGM and OD terms are very sensitive to small perturbations, especially for increasingly large unit cells. The significantly decreased PGM contribution can be mostly attributed to the reduced (diagonal) phonon group velocities. As shown in Fig.\ref{fig:disorder}C and D, the introduction of mass disorder strongly breaks the energy degeneracy of phonon bands and suppresses the (diagonal) group velocities, whereas phonon frequencies change by much less. This results in an increased OD contribution due to the lifting of degeneracy, which makes the otherwise vanishing off-diagonal velocities appreciable~\cite{Hardy1963}. We can also infer from Fig.\ref{fig:disorder}A that the PGM-\minikappa\ tends toward zero when $N_{\rm a}$ approaches infinity while OD-\minikappa\ approaching the total \minikappa. Interestingly, such a disorder-induced interconversion between PGM-\minikappa\ and OD-\minikappa, when approached from an inverse perspective, might be leveraged to estimate the total \minikappa\ without explicitly computing the OD contributions. That is, we may approximate the total \minikappa\ of some very complex materials (large $N_{\rm a}$) wherein OD-\minikappa\ dominates by computing only the PGM-\minikappa\ of a simplified model (e.g. $N_{\rm a}$ = 1). The latter could be obtained by averaging the atomic masses, positions, and IFCs of the complex structure. \minikappa\ based on such an approximation can be easily calculated using macroscopic properties such as materials' density and elastic properties.


We have so far qualitatively answered the two questions raised earlier by means of showing the impacts of weak disorder and structural complexity on \minikappa\ in an idealized model. However, the assumption of weak disorder makes the above picture applicable to only a small group of compounds which share similar structures, chemical compositions, and interatomic interactions. To better describe complex materials with large structural and compositional fluctuations, we move on to investigate models with increasingly strong disorder. We show in Fig.\ref{fig:disorder}B the calculated total and decomposed \minikappa\ as a function of the strength of random mass disorder (our tests show that perturbing IFCs has similar effects). We see from Fig.\ref{fig:disorder}B that PGM-\minikappa\ decreases sharply with the onset of the disorder and then continues to decrease in a slower manner with increasing disorder. In contrast, OD-\minikappa\ increases first and then decreases with enhanced disorder, while its percent contribution keeps increasing (see Supplementary Materials). The resulting total \minikappa\ displays an overall reduced value. By comparing the left and right panels in Fig.\ref{fig:disorder}B, we also observe that a larger unit cell with similar strength of disorder tends to shift the maximum of OD-\minikappa\ towards the weaker disorder regime and make OD-\minikappa\ more important. 

Overall, we see that, in the full range of disorder from weak to strong, the total \minikappa\ of complex unit cells might be approximated by computing only PGM-\minikappa\ for the perfectly ordered crystal, despite the latter displaying certain underestimation. Surprisingly, the presence of strong disorder might accidentally lead to a better agreement between the two, as can be inferred by comparing OD-\minikappa\ without any disorder and the total \minikappa\ with strong disorder in Fig.\ref{fig:disorder}B. These results help explain why we find that the total PGM+OD-\minikappa\ in Fig.\ref{fig:minikappa}B lies in a relatively narrow range, which is bounded both from above and below over a variety of compounds. It can be interpreted as follows: (i) complex compounds are nothing but variants of simple compounds with additional compositional and structural disorder; (ii) despite potentially dissimilar amounts of structural complexity, different compounds share close values of \minikappa\ if their averaged structural properties, such as mass, position, and interatomic interaction, are similar; (iii) it is likely that the averaged structural properties depend weakly on structural complexity for compounds with similar chemistries.

\begin{figure*}[htp]
	\centering
	\includegraphics[width = 1.0\linewidth]{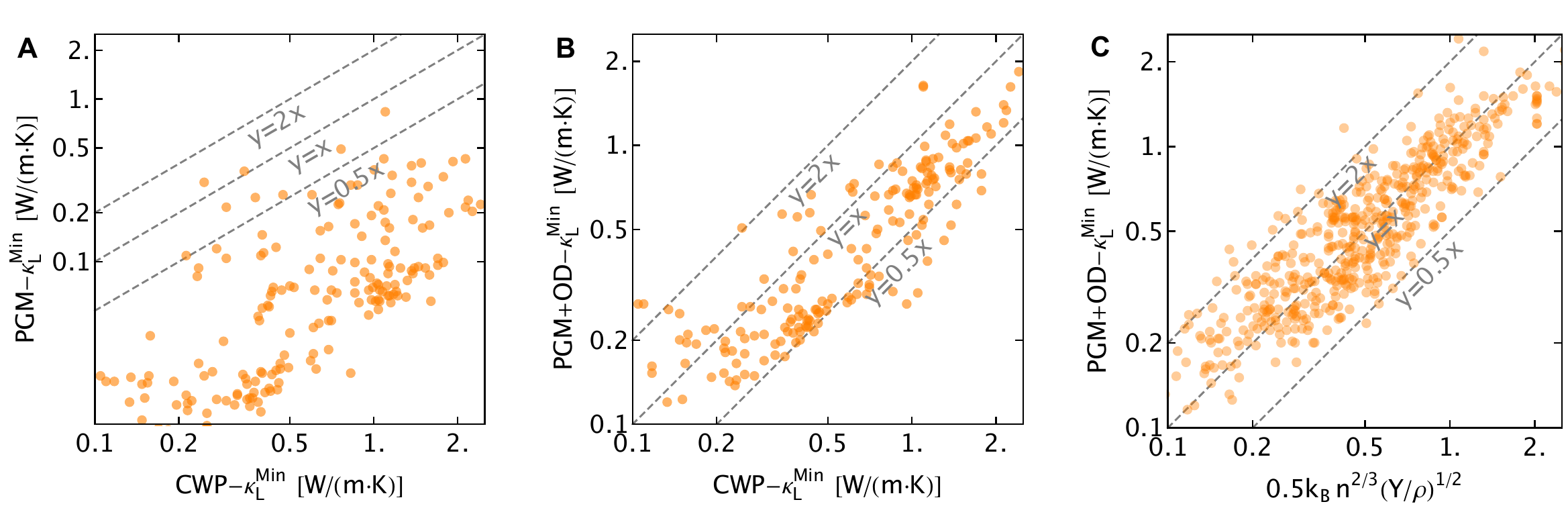}
	\caption{
	Comparisons of the minimum lattice thermal conductivity based on Eq.[\ref{eq:minikappa}] with the CWP model \textcolor{red}{for cubic crystals at 600~K.} (A) Comparisons of the phonon gas model of \minikappa\ (PGM-\minikappa) with the CWP model (CWP-\minikappa). (B)Comparisons of the unified minimum thermal transport model that combines both phonon gas model and the off-diagonal terms (PGM+OD-\minikappa) with the CWP model. (C) The linear correlation of the PGM+OD-\minikappa\ with an approximated formula of the CWP model in the high temperature limit, i.e., $0.5k_{\rm B} n^{2/3}(Y/\rho)^{1/2}$, wherein $n$, $Y$, and $\rho$ are the number density, Young's modulus, and mass density, respectively. The dashed lines in (A), (B), and (C) denote the deviations from the diagonal within a factor of two. 
	}
	\label{fig:cwp}
\end{figure*}


We apply these newly developed insights to bridge the knowledge gap between our \minikappa\ model and the CWP model. This is motivated by the missing connection between the CWP model and Allen and Feldman's theory of diffusons, a question initially raised by Cahill and Pohl~\cite{Cahill1989}. Considering the PGM nature of the CWP model, we first compare the PGM-\minikappa\ calculated using our model to those from the CWP model (denoted as CWP-\minikappa). To eliminate the uncertainty caused by an anisotropic average of \minikappa, we only selected compounds with cubic symmetry, as shown in Fig.\ref{fig:cwp}A. In contrast to the expected failure of the CWP model due to the lack of OD contribution, we find that CWP-\minikappa\ deviates significantly from PGM-\minikappa\ and displays much larger values, which seems to overcome the underestimation inherent in PGM-\minikappa. The latter is further confirmed by plotting CWP-\minikappa\ against PGM+OD-\minikappa\ in Fig.\ref{fig:cwp}B. Overall, we find that CWP-\minikappa\ is comparable to PGM+OD-\minikappa, although the latter tends to show smaller values statistically.

The above comparison reveals that the CWP model seems to work remarkably well, which is quite surprising because (i) the CWP model is relatively simple and only needs sound velocity and number density as inputs [see Eq(3) in Ref.\cite{Cahill1992}], and (ii) the CWP model does not at all account for the OD contribution, which is dominated by optical phonon modes~\cite{Simoncelli2019}. We attribute the success of the CWP model to the fact that the CWP model could be a good approximation to the sophisticated PGM+OD-\minikappa\ model based on Eq.(\ref{eq:minikappa}) by averaging out disorder and phonon unfolding, in the same spirit of our model detailed earlier. This is achieved by neglecting the details of both chemical composition and atomic arrangement using a single parameter, i.e., the number density $n$, to describe the structure in the CWP model. In such a simplified picture with only one atom in an averaged unit cell, only three acoustic branches arise, which induces an effective back conversion from the OD contribution to the PGM contribution, thus making the PGM dominant again. This picture is supported by Fig.\ref{fig:minikappa}C, which shows that PGM-\minikappa\ tends to dominate over OD-\minikappa\ in simple crystals. Importantly, the above analysis implies that it is inappropriate to use the CWP model as a second heat transport channel on top of the PGM model as adopted in recent studies~\cite{Mukhopadhyay1455,Minghui2019}, which will result in a double-counting of the PGM contribution.

The interpretation of the CWP model based on the structure averaging and phonon unfolding picture offers additional insights into its potential limitations or uncertainties. On one hand, neglecting structural details might lead to a lack of additional flattening of phonon dispersions caused by disorder or symmetry breaking, thus giving rise to an overestimated \minikappa. On the other hand, OD contributions from the three acoustic branches are still missing, which leads to a general underestimation of \minikappa. Considering these two competing factors, there is no clear answer on the net effect. However, through our numerical experiment presented in Fig.\ref{fig:cwp}B, we see that CWP-\minikappa\ is likely to overestimate \minikappa, thus indicating that the dominant uncertainty probably comes from the first factor. This might explain why experiments tend to find lower lattice thermal conductivities than the values predicted by the CWP model~\cite{Agne2018}. 


To establish a quantitative measure of the extent that the CWP model overestimates \minikappa\ compared to our model, in Fig.\ref{fig:cwp}C, we numerically fit our calculated PGM+OD-\minikappa\ based on the relation of $\kappa_{\rm L}^{\rm min} \propto k_{\rm B} n^{2/3} (Y/\rho)^{1/2}$ suggested by Clarke~\cite{CLARKE200367}, wherein $n$ is the number density, $Y$ is Young's modulus, and $\rho$ is the mass density. We find that PGM+OD-$\kappa_{\rm L}^{\rm min}  \approx 0.5k_{\rm B} n^{2/3} (Y/\rho)^{1/2}$, which is considerably smaller than that from the CWP model in the high temperature limit~\cite{Cahill1992}, i.e., $\kappa_{\rm L}^{\rm min}  \approx 1.1k_{\rm B} n^{2/3} (Y/\rho)^{1/2}$ if sound velocity is approximated as $v_{s} = (2v_{\rm T}+v_{\rm L})/3 \approx 0.94(Y/\rho)^{1/2}$~\cite{CLARKE200367,Agne2018}. Our analysis unambiguously demonstrates that the CWP model tends to overestimate \minikappa\ intrinsically, thus offering a different perspective from the proposal by Agne \textit{et al.}~\cite{Agne2018}, who propose to mitigate such an overestimation by resorting to an alternative heat transfer through diffusons. We note that our conclusion only applies to the general behavior of the CWP model but may not work for a specific compound, as indicated by the large spread of data points along the diagonal in Fig.\ref{fig:cwp}C. Furthermore, we see from the relation of $\kappa_{\rm L}^{\rm min} \propto k_{\rm B} n^{2/3} (Y/\rho)^{1/2}$ that the bound to \minikappa\ of crystals can be attributed to the competing parameters of $n$, $Y$, and $\rho$, all of which might have physical bounds for crystals.

\begin{figure*}[ht]
	\centering	
	\includegraphics[width = 1\linewidth]{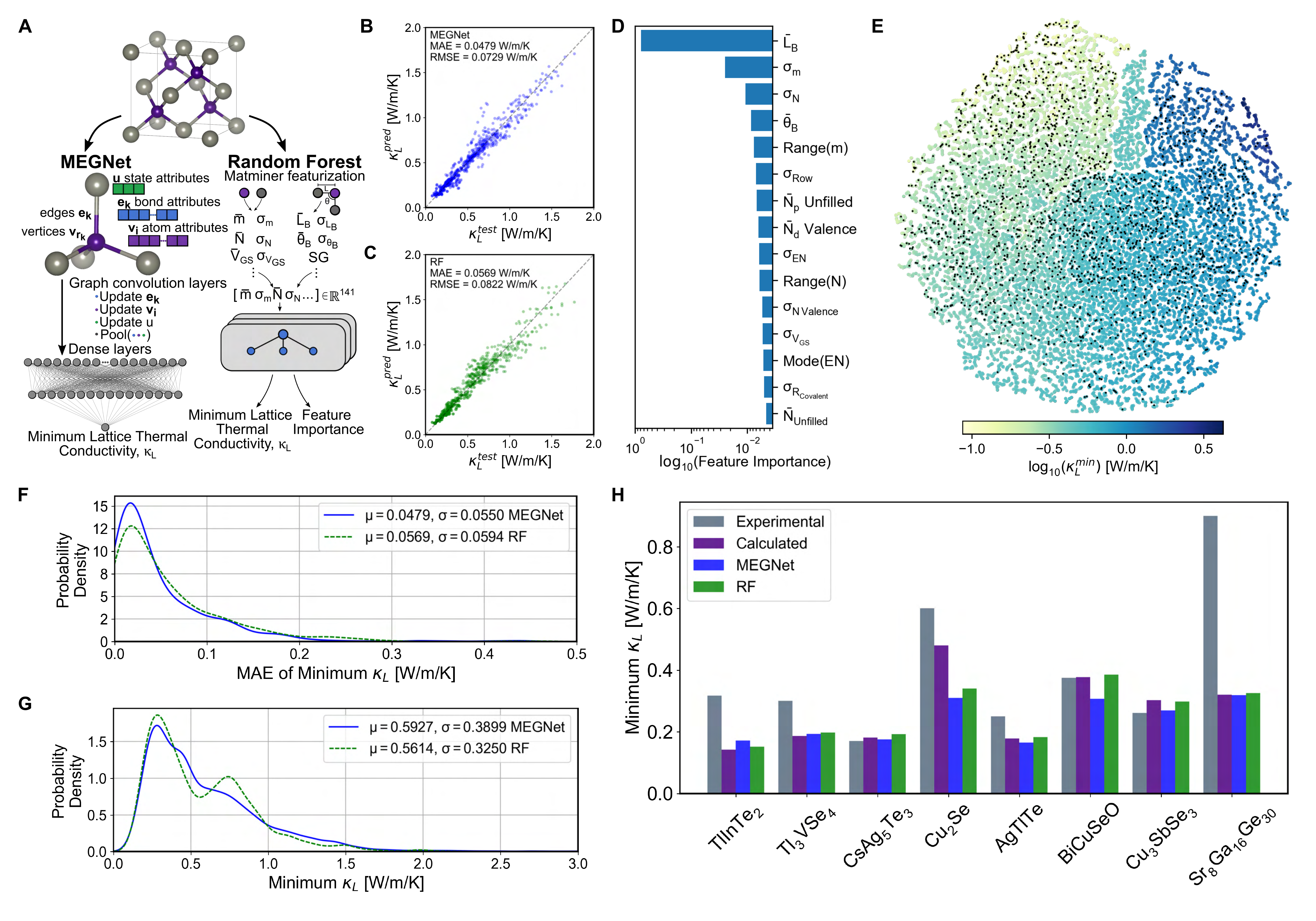}
	\caption{
	Machine learning models of minimum lattice thermal conductivity. (A) Depiction of the two machine-learning models, and the error on the test set for (B) MEGNet and (C) the random forest. (D) Feature importance from the random forest. (E) MEGNet latent graph features reduced to two dimension through t-SNE. Black points show the training set and other points show the ICSD dataset, colored by their estimated \minikappa. (F) Kernel density estimate of the MAE distribution for \minikappa\ for both models on the test set. (G) Kernel density estimate of \minikappa\ for both models on the ICSD dataset. (H) \minikappa\ of seven selected compounds from experiments (TlInTe$_{2}$ at 600~K~\cite{TlInTe2}, Tl$_{3}$VSe$_{4}$ at 300~K~\cite{Mukhopadhyay1455}, CsAg$_5$Te$_3$ at 600~K~\cite{CsAg5Te32016}, Cu$_2$Se at 600~K~\cite{Liu:2012aa}, AgTlTe at 600~K~\cite{AgTlTe2005}, BiCuSeO at 600~K~\cite{BiCuSeO}, Cu$_3$SbSe$_3$ at 600~K~\cite{Cu3SbSe3},  and Sr$_8$Ga$_{16}$Ge$_{30}$ at 300~K~\cite{SGGNolas}), direct calculation using our PGM+OD-\minikappa\ model, MEGNet, and the random forest at the corresponding temperatures. All \minikappa were isotropically averaged.
	}
	\label{fig:minikappaml}
\end{figure*}

With our \minikappa\ model established, we wanted to investigate the distribution of \minikappa\ across all experimentally known compounds in the Inorganic Crystal Structure Database (ICSD)~\cite{icsd}. Due to the high computational cost of calculating phonon properties from first-principles, it would be unfeasible to calculate them all, making it necessary to use a more efficient model. We decided on two complementary approaches to machine-learn \minikappa: MatErials Graph Network (MEGNet)~\cite{Chen2019} -- a state of the art crystal graph convolutional neural network -- and a random forest -- a more interpretable approach (as seen in Fig.\ref{fig:minikappaml}A). MEGNet directly uses crystal structure information to construct crystal graphs as features. In contrast, the random forest requires featurization. For the random forest, we used a variety of features generated by Matminer~\cite{matminer} using the Magpie~\cite{Ward:2016aa} preset including the mean, standard deviation, minimum, maximum, mode, and range of elemental properties such as atomic number ($N$), mass ($m$), electronegativity ($EN$), Mendeleev number ($N_m$), melting temperature ($T_m$), column, row, covalent radius ($R_{\rm covalent}$), number of electrons and unfilled slots in the $s$, $p$, $d$, and $f$ valence orbitals, bandgap ($E_{{g}_{\rm GS}}$), magnetic moment (M$_{\rm GS}$), and volume per atom in the ground state ($V_{\rm GS}$). We also included some simple structural features such as the space group, and the mean and standard deviations of both bond length ($L_{B}$) and angle ($\theta_{B}$), totaling 141 features for each compound. We refer the readers to the Methods section for more details on training machine learning models.

The two models achieve similar accuracies on our test set, with MEGNet slightly outperforming the random forest (Fig.\ref{fig:minikappaml}B,C, and F), and both also predict similar distributions for \minikappa\ on the ICSD data set (Fig.\ref{fig:minikappaml}G). The distribution is bounded on the low end near 0.1~W/[m$\cdot$K] and a maximum around 4~W/[m$\cdot$K]. Only $\sim$100 of the $\sim$36,000 structures show MEGNet-predicted \minikappa\ values above 2.64 W/[m$\cdot$K] (the maximum observed in the calculated training set) and are almost entirely various allotropes of Be, C, and Co or carbides and nitrides, all having strong bonding and relatively low densities. Part of the difference in predicted values between the models arises from a limitation of the random forest, as it is restricted to interpolating between closest data points in our training set, preventing it from predicting smaller or larger values than those it was trained on. On the other hand, MEGNet is capable of extrapolation and finds more compounds with higher values of \minikappa\ when compared to the random forest and finds a smoother distribution for intermediate values, as shown in Fig.\ref{fig:minikappaml}G.

The random forest model provides advantages in the form of faster computation, interpretable feature importance, and easily visualized features. From the random forest, we extracted the 15 most important features and have plotted them in Fig.\ref{fig:minikappaml}D. Our most information-dense features are related to bonding strength ($L_{\rm B}$, $\theta_{\rm B}$, $EN$, $N_{\rm valence}$) and factors concerning the atomic properties ($m$, $N$, $\rho$, $R_{\rm covalent}$, $V_{\rm GS}$). In some cases, the mean properties are more useful, such as for bond length ($L_{B}$), bond angle ($\theta_{\rm B}$), and valence electron counts ($N_{\rm valence}$), while in other cases, the differences between atomic properties carry more information ($\sigma_{m}$,  $\sigma_{N}$, $\sigma_{Row}$, $\sigma_{EN}$). This is not unexpected, as phonon transport is sensitive to both similarities and differences within compounds. For example, we tend to expect shorter bonds to be stronger which results in higher characteristic frequencies, while mass difference among atoms could significantly alter the phonon spectrum. Notably, the most information-dense features being related to bonding strength is consistent with the positive correlation with $Y$ in the relation of $\kappa_{\rm L}^{\rm min} \propto k_{\rm B} n^{2/3} (Y/\rho)^{1/2}$.

Finally, we performed dimensionality reduction using t-SNE (t-Distributed Stochastic Neighbor Embedding) with PCA (Principal Component Analysis) initialization in order to project MEGNet's latent graph features into two dimensions for visualization, as seen in Fig.\ref{fig:minikappaml}E. The distribution of the training set and the ICSD dataset are considered together, and the overlap generally tells us how well the training set covers the ICSD dataset, indicating potential transferability of the model. The points containing ICSD materials with low and high predicted \minikappa\ are segregated in the latent space, allowing the model to clearly separate them from each other.  Additionally, the training set points span the extent of the axes and their distribution varies smoothly, with the exception of less density near some outer edges and a small cluster towards the top center of the figure. Nevertheless, the coverage is especially dense in the area of low predicted \minikappa\ which is most relevant for thermoelectric or thermal barrier coating materials, suggesting that our training set sufficiently covers the space, but coverage could be improved by increasing the size of our training set in the future.


We further validate these predictions against experimental measurements and direct calculations using Eq.$[\ref{eq:minikappa}]$ Since \minikappa\ is relatively rare in ordered crystals, we deliberately choose thermoelectric materials exhibiting ultralow $\kappa_{\rm L}$, which presumably best represent \minikappa, as shown in Fig.\ref{fig:minikappaml}H. We find that the experimental values are either higher or comparable to PGM+OD-\minikappa, consistent with the definition of \minikappa. It is noteworthy that compounds including CsAg$_{5}$Te$_{3}$~\cite{CsAg5Te32016}, Cu$_3$SbSe$_3$~\cite{Cu3SbSe3}, and BiCuSeO~\cite{BiCuSeO} have $\kappa_{L}$ already close to \minikappa\ from our model, suggesting that further improvement in thermoelectric efficiency should be focused on optimizing their electronic transport properties. We note that CsAg$_{5}$Te$_{3}$ and Cu$_3$SbSe$_3$ indeed resemble amorphous solids and display nearly temperature-independent and glasslike $\kappa_{\rm L}$~\cite{CsAg5Te32016,Cu3SbSe3}. Whereas for compounds such as Tl$_{3}$VSe$_{4}$~\cite{Mukhopadhyay1455} and TlInTe$_{2}$~\cite{TlInTe2}, there is still room for further reducing $\kappa_{\rm L}$, although their $\kappa_{\rm L}$ are already very low. Also, we find it interesting to  see a large gap between our calculated \minikappa\ and the experimentally measured $\kappa_{\rm L}$ in Sr$_8$Ga$_{16}$Ge$_{30}$\cite{SGGNolas,CHAKOUMAKOS200080}, a typical electron-crystal phonon-glass material. This finding implies that although $\kappa_{\rm L}$ of a material might exhibit a temperature dependence similar to that of glasses, $\kappa_{\rm L}$ may still not achieve the minimum.

Overall, the machine-learned \minikappa\ from both MEGNet and the random forest agree very well with our direct calculations. This is rather encouraging because most of the compounds are not contained in the training dataset for the machine learning models except for TlInTe$_{2}$. The largest deviation between experiment, machine learning models, and direct calculation is found in Cu$_{2}$Se. The discrepancy might be attributed to the following factors: (i) we used an ordered structure ($\beta$-Cu$_{2}$Se) to compute \minikappa, while in the experimental structure the copper ions are highly disordered around the Se sublattice~\cite{Liu:2012aa}; (ii) $\beta$-Cu$_{2}$Se often exhibits off-stoichiometry with copper vacancies, resulting in relatively large variations in the experimentally measured values of $\kappa_{\rm L}$ in the range of 0.4-0.6~W/[m$\cdot$K] among different samples~\cite{Liu:2012aa}. Therefore, despite the disordered nature of $\beta$-Cu$_{2}$Se, our models, including both machine learning and direct calculation, seem to provide reasonable estimations using ordered structures.

\begin{figure*}[ht]
	\centering	
	\includegraphics[width = 1.0\linewidth]{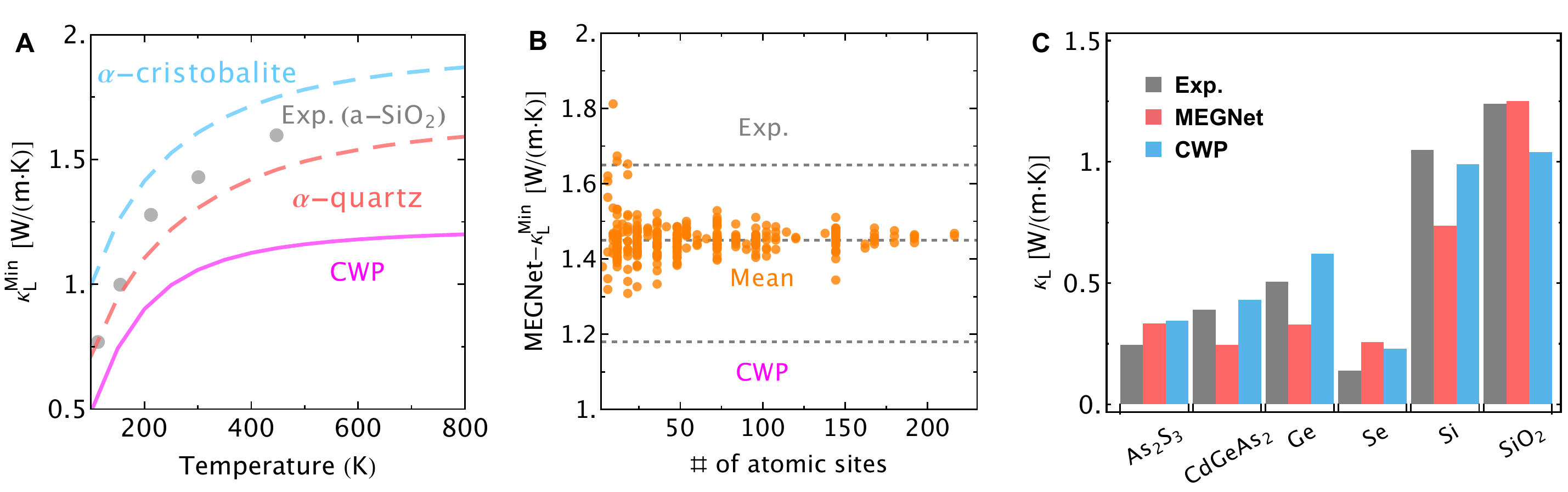}
	\caption{
	Prediction of lattice thermal conductivity in amorphous solids using machine learning models of minimimum lattice thermal conductivity. (A) Comparison of temperature-dependent $\kappa_{\rm L}$ of amorphous silica (a-SiO$_{2}$) between the experiment (gray disks)~\cite{Cahill1992} and theoretical \minikappa\ (solid/dashed lines). Results from the CWP model are indicated by solid magenta lines. The dashed lines display the calculated PGM+OD-\minikappa\ for two ordered phases of SiO$_{2}$, namely $\alpha$-quartz and $\alpha$-cristobalite. (B) Predicted \minikappa\ (orange disks) at 600~K using MEGNet for 317 structures of ordered SiO$_{2}$ from Materials Project~\cite{matproj} as a function of the number of atomic sites in a primitive cell. The upper, middle, and lower dashed lines denote the values from the experiment, mean of MEGNet predictions, and the CWP model, respectively. (C) Comparison of $\kappa_{\rm L}$ at 300~K of six amorphous compounds between experiments~\cite{Cahill1992}, our MEGNet model, and the CWP model~\cite{Cahill1992}.
	}
	\label{fig:amorphous}
\end{figure*}

The above results on Cu$_{2}$Se indicate that $\kappa_{\rm L}$ of amorphous compounds might be approximated with a  reasonable accuracy using ordered crystals. To confirm this hypothesis, we further apply our model to amorphous silica (a-SiO$_{2}$). We calculated PGM+OD-\minikappa\ for two ordered phases of SiO$_{2}$ ($\alpha$-quartz and $\alpha$-cristobalite) that are free of lattice instabilities, which are compared to the CWP model and the experimental measurement~\cite{Cahill1992} in Fig.\ref{fig:amorphous}A. We see that PGM+OD-\minikappa\ of both $\alpha$-quartz and $\alpha$-cristobalite agree well with experiment, displaying overall improvement over the CWP model, particularly at high temperatures. This implies that specific atomic arrangement might not be very relevant when approaching \minikappa, although it is responsible for subtle differences, which are probably more profound at low temperatures.

To provide more evidence for this observation, we next examine other phases of SiO$_{2}$ using our machine learning model. Particularly, we used MEGNet instead of the random forest because the former better encodes structural information. Since our MEGNet model is trained on structures from the Materials Project (MP)~\cite{matproj,pymatgen,mpapi}, we collected all structures with the chemical formula SiO$_{2}$ available within MP, totaling 317 structures. The calculated \minikappa\ using MEGNet is shown in Fig.\ref{fig:amorphous}B. We find a relatively large spread of \minikappa\ when $N_{a}$ is small ($N_{a} < 25$), while the spread becomes much smaller when $N_{a}$ is large ($N_{a} > 75$). The mean value of \minikappa\ of the 317 structures is 1.45~W/[m$\cdot$K], close to the experimental value of 1.65~W/[m$\cdot$K] for the amorphous phase~\cite{Cahill1992} and better than the value of 1.18~W/[m$\cdot$K] from the CWP model at 600~K~\cite{Cahill1992}. These results confirm our earlier hypothesis that $\kappa_{\rm L}$ of amorphous compounds can be approximated by ordered crystals. Moreover, a reliable prediction may be achieved by averaging over different structures, effectively capturing the distinct local structures in amorphous solids.

We apply our MEGNet model to other amorphous solids by averaging over ordered phases and compared to the experiments and the CWP model in Fig.\ref{fig:amorphous}C. Our machine learning model achieves an accuracy comparable with the CWP model. The advantage of our model is the calculation requires very low computational cost (within a second for each compound) and simple input (only structural information), and thus is readily applicable to many other compounds on a large scale. We note that compared to the amorphous SiO$_{2}$ our predicted $\kappa_{\rm L}$ of amorphous Si is considerably smaller than both the CWP model and the experiment. We attribute such a discrepancy mainly to the observations reported in the literature that (i) propagating modes contribute significantly more in amorphous Si than amorphous SiO$_{2}$, and (ii) phonon lifetimes in amorphous Si are notably larger than $\pi/\omega$~\cite{PhysRevB.89.144303}, both of which indicate that $\kappa_{\rm L}$ in amorphous Si still does not reach the minimum value. We hypothesize that this might also be the case for amorphous CdGeAs$_2$ and Ge.

Before closing, we briefly comment on the limitations of our theoretical and machine learning models. Firstly, both the identified lower bound to \minikappa\ and the constructed machine learning models are based on the isotropically averaged \minikappa. Consequently, it is reasonable to question how \minikappa\ behaves in anisotropic structures. This anisotropy is of particular interest because $\kappa_{\rm L}$ as small as 0.05 W/[m$\cdot$K] below our isotropically averaged lower bound at room temperature has been realized in layered WSe$_2$ crystals~\cite{chiritescu2007ultralow}. Therefore, we analyzed the anisotropic \minikappa\ for all of the calculated compounds (see Fig.S3 in Supplementary Materials). We find there are 67 compounds in total that have \minikappa\ less than 0.1 W/[m$\cdot$K] in at least one of the three Cartesian directions. However, there are only 5 compounds exhibiting \minikappa\ less than 0.07 W/[m$\cdot$K] (see listed compounds and \minikappa\ in Supplementary Materials). This observation suggests that, even though it may be difficult to find such materials due to their rarity, anisotropic materials, including layered crystals, have the potential to achieve a lattice thermal conductivity of less than 0.1 W/[m$\cdot$K], consistent with the prior work~\cite{chiritescu2007ultralow}.

Secondly, we computed all \minikappa\ using the harmonic phonon dispersion in the absence of finite temperature induced phonon frequency shifts; the latter are important in a range of low-thermal-conductivity materials~\cite{tvs2020,rczb2020}. As a first attempt to investigate the effects of phonon anharmonicity on \minikappa, we computed \minikappa\ at 300 K for Tl$_3$VSe$_4$ using harmonic and anharmonic phonon dispersions, respectively. The anharmonic phonon dispersion was computed using the self-consistent phonon theory at 300~K in a previous study~\cite{tvs2020}. We find that accounting for anharmonic effects only leads to a small change (about 7\% decrease) in \minikappa, which is much smaller than the change (about 162.5\% increase) in $\kappa_{\rm L}$ when phonon lifetimes are limited by intrinsic three phonon scattering~\cite{tvs2020}. The relative insensitivity of \minikappa\ to anharmonic effects can be explained by the relationship $\kappa_{\rm L}^{\rm min} \propto k_{\rm B} n^{2/3} (Y/\rho)^{1/2} \propto k_{\rm B} n^{2/3} v_s $~\cite{CLARKE200367}, which indicates that \minikappa\ is largely unchanged unless anharmonic effects lead to a  sizable change in sound velocity $v_s$.

Thirdly, it is not entirely justified to use one half of the vibrational period as the minimum phonon lifetime, i.e., $\tau = \pi/\omega$. The adoption of $\tau = \pi/\omega$ was first proposed by Cahill, Watson, and Pohl~\cite{Cahill1992}, motivated by Einstein's thermal conductivity model~\cite{Einstein1911}. Specifically, Einstein's model relies on a critical assumption that there is no coherence between the motions of neighboring atoms, which seems to be a reasonable choice for glassy materials. Einstein further derived the energy exchange between neighboring atoms by integrating over one half of a period of oscillation, however, without an explicit explanation in his original paper~\cite{Einstein1911}. We hypothesize this might be motivated by the fact that it is the shortest time period that an atom returns to its original position after exchanging energy with neighboring atoms. Cahill \textit{et al.} interpreted Einstein's result to mean that each atom undergoes a thermal energy loss or gain in the span of half an oscillation period~\cite{Cahill1992} and subsequently assumed that $\tau = \pi/\omega$. Mathematically, $\tau$ adopted in computing \minikappa\ can have an arbitrary value, which presumably is small in order to reduce the PGM contribution. However, physically, $\tau$ cannot be arbitrarily short  because our derived equation for \minikappa\ relies on a critical assumption that phonons are well-behaved quasiparticles ($\tau\omega \approx 2\pi$). The very small $\tau$ implies strong interactions and will invalidate the quasiparticle picture, making the phonon properties (such as group velocities and frequencies) less well-defined and invalidating the usage of Eq.[1]. To explore the potential uncertainty due to the adopted $\tau$, we performed sensitivity analysis of \minikappa\ by varying $\tau$ for a randomly selected set of compounds (see Supplementary Materials). We find that \minikappa\ decreases for most compounds when $\tau$ is decreased. Specifically, a sharp decrease in \minikappa\ is found when $\tau$ is reduced from a larger value to $\pi/\omega$, a value we adopted for computing \minikappa. When $\tau$ is less than $\pi/\omega$, slower decrease or no change in \minikappa\ is observed, especially for lower values of \minikappa. Since $\tau$ cannot be arbitrarily small as argued previously, we hypothesize that $\tau = \pi/\omega$ might be a reasonable choice to achieve a relatively low value of \minikappa\ while approaching the limit of the quasiparticle picture.

Finally, despite encouraging results from the comparison between the calculations and the experiments, our theoretical model only accounts for harmonic heat flux~\cite{Hardy1963,Sun2010} and Eq.[\ref{eq:minikappa}] fails to describe phonon satellite structures deviating from the quasiparticle picture~\cite{delaire2011giant}. For the machine learning model, local structural motifs may not be well captured by ordered structures for amorphous solids, and our simple averaging scheme with equal weight for each structure is probably not optimal. We deem establishing a more reliable physical bound to phonon lifetime, including higher-order anharmonic effects on the phonon frequency shifts and the heat flux beyond the phonon quasiparticle picture, and machine learning structural motifs inherent in amorphous solids are interesting avenues of further research to improve our model.

\vspace{1.0cm}
\section{CONCLUSIONS}
In summary, we have developed a first-principles minimum lattice thermal conductivity model based on a unified theory of thermal transport in crystals and glasses. By applying such a model to thousands of crystalline compounds, we have discovered a universal bound to minimum lattice thermal conductivity independent of structural complexity. In striking contrast to the conventional phonon gas model, such an unusual behavior is found to be deeply rooted in the conversion of the heat transfer mechanism from the phonon gas model to the diffuson picture with the presence of disorder, while the value of the total minimum lattice thermal conductivity largely remains unchanged. With these insights, we bridge the knowledge gap between the Cahill-Watson-Pohl model and our unified model by pointing out that the former could be viewed as an approximation of the latter. We further construct machine learning models based on a graph network and random forest to enable fast and accurate prediction of minimum lattice thermal conductivity on a large scale, which is validated against thermoelectric materials with ultralow and glasslike lattice thermal conductivity. We demonstrate the applicability of our graph network model to simulate lattice thermal conductivity in amorphous solids. These findings highlight a unified understanding of the lower limit of lattice thermal transport in solids. The first-principles-based theory and machine learning model built in this work are universal and readily applicable in research relevant to thermal energy conversion and management.

\begin{acknowledgments}
Y.X., D.G., K.P., and C.W. acknowledge financial support received from (i) Toyota Research Institute (TRI) through the Accelerated Materials Design and Discovery program (thermal conductivity calculations), (ii) the Department of Energy, Office of Science, Basic Energy Sciences under grant DE-SC0014520 (theory of anharmonic phonons), and (iii) the U.S. Department of Commerce and National Institute of Standards and Technology as part of the Center for Hierarchical Materials Design (CHiMaD) under award no. 70NANB14H012 (DFT calculations). Y. X. was also supported by Portland State University Lab Setup Fund. M.G.K. were supported in part by the National Science Foundation Grant DMR-2003476. V.O. acknowledges financial support from the National Science Foundation Grant DMR-1611507. We acknowledge the computing resources provided by (i) the National Energy Research Scientific Computing Center (NERSC), a U.S. Department of Energy Office of Science User Facility operated under Contract No. DEAC02-05CH11231, (ii) Quest high-performance computing facility at Northwestern University which is jointly supported by the Office of the Provost, the Office for Research, and Northwestern University Information Technology, and (iii) Bridges2 at Pittsburgh Supercomputing Center (PSC) through allocations dmr160027p and mat220007p from the Advanced Cyber-infrastructure Coordination Ecosystem: Services \& Support (ACCESS) program, which is supported by National Science Foundation grants \#2138259, \#2138286, \#2138307, \#2137603, and \#2138296. The authors (Y.X., C.W, and M.K, initial DFT and conception of research problem) also acknowledge support for the initial stages of this research from the U.S. Department of Energy under Contract No. DE-SC0014520. \end{acknowledgments}
\appendix


\section{First-principles calculation of \minikappa}

The key ingredients for modeling \minikappa\ based on Eq.[\ref{eq:minikappa}] from first principles are materials harmonic vibrational spectra, which can be explicitly calculated if materials structure and harmonic interatomic force constants are known. To construct a large database for \minikappa, we used the phonon database (phononDB) generated by Togo~\cite{phonondbtogo}, who used crystalline structures from the Materials Project (MP)~\cite{matproj,pymatgen,mpapi} and calculated harmonic interactions using Phonopy~\cite{Togo2015}. It is worth noting that these phonon calculations were performed by means of the Vienna {\it Ab\ Initio\/} Simulation Package (VASP)~\cite{Vasp1, Vasp2, Vasp3, Vasp4}, which employed the projector-augmented wave (PAW)~\cite{paw} method in conjunction with the revised Perdew-Burke-Ernzerhof version (PBEsol~\cite{Perdew2008}) of the generalized gradient approximation (GGA)~\cite{gga} for the exchange-correlation functional~\cite{dft}. We performed post processing of the phononDB (version of 2018-04-17) to generate harmonic force constants and downselected the compounds that are free of lattice instabilities (imaginary phonon frequencies) with supercell structures constructed from diagonal matrices (required by our \minikappa\ implementation within ShengBTE~\cite{shengbte}), totaling 2576 compounds. These compounds cover wide ranges of chemical compositions and space group symmetries, with 189, 922, 310, 660, and 495 compounds from cubic, orthorhombic, tetragonal, trigonal/hexagonal, triclinic/monoclinic crystal systems, respectively. Other compounds and their harmonic phonon properties, including TlInTe$_{2}$, Tl$_{3}$VSe$_{4}$, CsAg$_{5}$Te$_{3}$, Cu$_{2}$Se, TlAgTe, CuBiSeO, SnSe, $\alpha$-cristobalite, and $\alpha$-quartz, were calculated following the similar DFT settings. For $\beta$-Cu$_{2}$Se which has imaginary phonon frequencies, we have performed anharmonic phonon renormalization at finite temperature (600~K) using self-consistent phonon theory~\cite{Errea2014,Tadano2015,rczb2020,tvs2020}.

We implemented the \minikappa\ model based on Eq.[\ref{eq:minikappa}] within ShengBTE~\cite{shengbte}. The off-diagonal group velocities were calculated following the derivations by Allen and Feldman~\cite{Allen1993disorder} and Simoncelli \textit{et al.}~\cite{Simoncelli2019}. We calculated \minikappa\ for these 2576 compounds at three temperatures of 300~K, 600~K, and 900~K using a constant mesh density (mesh\_density = 50) following Phonopy conventions~\cite{Togo2015}. The convergence is carefully monitored across different compounds to achieve good balance between accuracy and efficiency.

\section{Machine learning models}

MEGNet: For the MEGNet model, we used frozen initial atom embeddings transferred from another MEGNet model~\cite{Chen2019} trained on 133,420 formation energies from Materials Project~\cite{matproj}. Crystal graphs were constructed using 50 features per bond with a gaussian smearing width of 0.5~\r{A} up to a cutoff radius of 5~\r{A} for the bond attributes, transferred atom embedding weights of size 16 for the atom attributes, and one single global attribute embedding the temperature in Kelvin. For the model hyperparameters, we used only a single MEGNet block and set the number of dense layers in the MEGNet block to $n_1$ = 16, $n_2$ = 16, and $n_3$ = 8, followed by 2 Set2Set passes, and the learning rate was set to 10$^{-3}$. In order to prevent overfitting, we set weight dropout to 25\% during both training and prediction and added a small L2 weight regularization parameter of 10$^{-5}$. The \minikappa\ data was split 80\%:10\%:10\% into a training set, validation set, and testing set. Predictions were then run over 10 trials and the average was taken as the final output. The model was run for 500 epochs with a patience of 250, and the model with the best validation performance was selected for further predictions. The MEGNet training mean absolute error (MAE) was 0.033~W/[m$\cdot$K], the validation MAE was 0.0406~W/[m$\cdot$K], and the testing MAE was 0.0479~W/[m$\cdot$K]. Finally, we performed dimensionality reduction on the combined phononDB and ICSD dataset using t-SNE to project the latent graph features into 2 dimensions. We intialized the t-SNE with PCA and chose the settings of 1,000 iterations, a perplexity of 5, and a learning rate of 39,514 (n).

Random Forest: Features for the random forest were created through Matminer~\cite{matminer} using the Magpie preset~\cite{Ward:2016aa}, along with global symmetry features and site statistic fingerprints, totaling 141 features for each compound as detailed in the main text. The random forest was constructed using 1,000 estimators, and we verified that including more estimators did not improve the accuracy significantly.  For this model, we used 5-fold cross validation on the 90\% training set for parameter selection and left a holdout test set of 10\%.  In order to prevent overfitting, we changed the allowable maximum depth of each tree, where larger maximum depths showed better prediction accuracies at the cost of a larger gap between training and testing accuracies. A maximum depth 8 was then chosen to balance training, validation, and testing accuracy. For the random forest, the mean training MAE was 0.0375~W/[m$\cdot$K], mean validation MAE was 0.0458~W/[m$\cdot$K], and the testing MAE was 0.0570~W/[m$\cdot$K]. Feature importance was extracted from a random forest trained on only 600 K data.

The training dataset used in both MEGNet and random forest models consists of our calculated \minikappa\ at three temperatures of 300K, 600K, and 900K, totaling 7728 datapoints. For the ICSD dataset, 36,199 crystal structures were pulled from the Open Quantum Materials Database (OQMD)~\cite{Saal:2013aa,Kirklin:2015aa}, and all machine learning predictions of \minikappa\ were then performed at 600 K unless otherwise specified.

\section{Data availability}
The codes, data sets, and machine learning models are available via public repository (\url{https://github.com/yimavxia/Minikappa})

\bibliography{minikappa}

\begin{thebibliography}{68}%
\makeatletter
\providecommand \@ifxundefined [1]{%
 \@ifx{#1\undefined}
}%
\providecommand \@ifnum [1]{%
 \ifnum #1\expandafter \@firstoftwo
 \else \expandafter \@secondoftwo
 \fi
}%
\providecommand \@ifx [1]{%
 \ifx #1\expandafter \@firstoftwo
 \else \expandafter \@secondoftwo
 \fi
}%
\providecommand \natexlab [1]{#1}%
\providecommand \enquote  [1]{``#1''}%
\providecommand \bibnamefont  [1]{#1}%
\providecommand \bibfnamefont [1]{#1}%
\providecommand \citenamefont [1]{#1}%
\providecommand \href@noop [0]{\@secondoftwo}%
\providecommand \href [0]{\begingroup \@sanitize@url \@href}%
\providecommand \@href[1]{\@@startlink{#1}\@@href}%
\providecommand \@@href[1]{\endgroup#1\@@endlink}%
\providecommand \@sanitize@url [0]{\catcode `\\12\catcode `\$12\catcode
  `\&12\catcode `\#12\catcode `\^12\catcode `\_12\catcode `\%12\relax}%
\providecommand \@@startlink[1]{}%
\providecommand \@@endlink[0]{}%
\providecommand \url  [0]{\begingroup\@sanitize@url \@url }%
\providecommand \@url [1]{\endgroup\@href {#1}{\urlprefix }}%
\providecommand \urlprefix  [0]{URL }%
\providecommand \Eprint [0]{\href }%
\providecommand \doibase [0]{http://dx.doi.org/}%
\providecommand \selectlanguage [0]{\@gobble}%
\providecommand \bibinfo  [0]{\@secondoftwo}%
\providecommand \bibfield  [0]{\@secondoftwo}%
\providecommand \translation [1]{[#1]}%
\providecommand \BibitemOpen [0]{}%
\providecommand \bibitemStop [0]{}%
\providecommand \bibitemNoStop [0]{.\EOS\space}%
\providecommand \EOS [0]{\spacefactor3000\relax}%
\providecommand \BibitemShut  [1]{\csname bibitem#1\endcsname}%
\let\auto@bib@innerbib\@empty
\bibitem [{\citenamefont {Bell}(2008)}]{Bell1457}%
  \BibitemOpen
  \bibfield  {author} {\bibinfo {author} {\bibfnamefont {Lon~E.}\ \bibnamefont
  {Bell}},\ }\bibfield  {title} {Cooling, Heating, Generating Power, and
  Recovering Waste Heat with Thermoelectric Systems,\ }\href {\doibase
  10.1126/science.1158899} {\bibfield  {journal} {\bibinfo  {journal}
  {Science}\ }\textbf {\bibinfo {volume} {321}},\ \bibinfo {pages} {1457--1461}
  (\bibinfo {year} {2008})}\BibitemShut {NoStop}%
\bibitem [{\citenamefont {Slack}(1979{\natexlab{a}})}]{slackbook2}%
  \BibitemOpen
  \bibfield  {author} {\bibinfo {author} {\bibfnamefont {Glen~A}\ \bibnamefont
  {Slack}},\ }\bibfield  {title} {The thermal conductivity of nonmetallic
  crystals,\ }\href@noop {} {\bibfield  {journal} {\bibinfo  {journal} {Solid
  state physics}\ }\textbf {\bibinfo {volume} {34}},\ \bibinfo {pages} {1--71}
  (\bibinfo {year} {1979}{\natexlab{a}})}\BibitemShut {NoStop}%
\bibitem [{\citenamefont {Slack}(1995)}]{slackbook}%
  \BibitemOpen
  \bibfield  {author} {\bibinfo {author} {\bibfnamefont {G.~A.}\ \bibnamefont
  {Slack}},\ }\href@noop {} {\emph {\bibinfo {title} {CRC Handbook of
  Thermoelectrics}}}\ (\bibinfo  {publisher} {CRC Press},\ \bibinfo {address}
  {Boca Raton, FL},\ \bibinfo {year} {1995})\ pp.\ \bibinfo {pages}
  {407--440}\BibitemShut {NoStop}%
\bibitem [{\citenamefont {Snyder}\ and\ \citenamefont
  {Toberer}(2008)}]{Snyder2008}%
  \BibitemOpen
  \bibfield  {author} {\bibinfo {author} {\bibfnamefont {G.~Jeffrey}\
  \bibnamefont {Snyder}}\ and\ \bibinfo {author} {\bibfnamefont {Eric~S.}\
  \bibnamefont {Toberer}},\ }\bibfield  {title} {Complex thermoelectric
  materials,\ }\href {http://dx.doi.org/10.1038/nmat2090} {\bibfield  {journal}
  {\bibinfo  {journal} {Nat. Mater.}\ }\textbf {\bibinfo {volume} {7}},\
  \bibinfo {pages} {105--114} (\bibinfo {year} {2008})}\BibitemShut {NoStop}%
\bibitem [{\citenamefont {Sootsman}\ \emph {et~al.}(2009)\citenamefont
  {Sootsman}, \citenamefont {Chung},\ and\ \citenamefont
  {Kanatzidis}}]{Sootsman2009}%
  \BibitemOpen
  \bibfield  {author} {\bibinfo {author} {\bibfnamefont {Joseph~R}\
  \bibnamefont {Sootsman}}, \bibinfo {author} {\bibfnamefont {Duck~Young}\
  \bibnamefont {Chung}}, \ and\ \bibinfo {author} {\bibfnamefont {Mercouri~G}\
  \bibnamefont {Kanatzidis}},\ }\bibfield  {title} {New and old concepts in
  thermoelectric materials,\ }\href@noop {} {\bibfield  {journal} {\bibinfo
  {journal} {Angewandte Chemie International Edition}\ }\textbf {\bibinfo
  {volume} {48}},\ \bibinfo {pages} {8616--8639} (\bibinfo {year}
  {2009})}\BibitemShut {NoStop}%
\bibitem [{\citenamefont {He}\ and\ \citenamefont {Tritt}(2017)}]{HeJian2017}%
  \BibitemOpen
  \bibfield  {author} {\bibinfo {author} {\bibfnamefont {Jian}\ \bibnamefont
  {He}}\ and\ \bibinfo {author} {\bibfnamefont {Terry~M.}\ \bibnamefont
  {Tritt}},\ }\bibfield  {title} {Advances in thermoelectric materials
  research: Looking back and moving forward,\ }\href {\doibase
  10.1126/science.aak9997} {\bibfield  {journal} {\bibinfo  {journal}
  {Science}\ }\textbf {\bibinfo {volume} {357}} (\bibinfo {year} {2017}),\
  10.1126/science.aak9997}\BibitemShut {NoStop}%
\bibitem [{\citenamefont {Clarke}(2003)}]{CLARKE200367}%
  \BibitemOpen
  \bibfield  {author} {\bibinfo {author} {\bibfnamefont {David~R.}\
  \bibnamefont {Clarke}},\ }\bibfield  {title} {Materials selection guidelines
  for low thermal conductivity thermal barrier coatings,\ }\href {\doibase
  https://doi.org/10.1016/S0257-8972(02)00593-5} {\bibfield  {journal}
  {\bibinfo  {journal} {Surface and Coatings Technology}\ }\textbf {\bibinfo
  {volume} {163-164}},\ \bibinfo {pages} {67--74} (\bibinfo {year} {2003})},\
  \bibinfo {note} {proceedings of the 29th International conference on
  Metallurgical Coatings and Thin Films}\BibitemShut {NoStop}%
\bibitem [{\citenamefont {Thakare}\ \emph {et~al.}(2021)\citenamefont
  {Thakare}, \citenamefont {Pandey}, \citenamefont {Mahapatra},\ and\
  \citenamefont {Mulik}}]{Thakare:2020aa}%
  \BibitemOpen
  \bibfield  {author} {\bibinfo {author} {\bibfnamefont {Jayant~Gopal}\
  \bibnamefont {Thakare}}, \bibinfo {author} {\bibfnamefont {Chandan}\
  \bibnamefont {Pandey}}, \bibinfo {author} {\bibfnamefont {MM}~\bibnamefont
  {Mahapatra}}, \ and\ \bibinfo {author} {\bibfnamefont {Rahul~S}\ \bibnamefont
  {Mulik}},\ }\bibfield  {title} {Thermal barrier coatings—A state of the art
  review,\ }\href@noop {} {\bibfield  {journal} {\bibinfo  {journal} {Metals
  and Materials International}\ }\textbf {\bibinfo {volume} {27}},\ \bibinfo
  {pages} {1947--1968} (\bibinfo {year} {2021})}\BibitemShut {NoStop}%
\bibitem [{\citenamefont {Cahill}\ \emph {et~al.}(1992)\citenamefont {Cahill},
  \citenamefont {Watson},\ and\ \citenamefont {Pohl}}]{Cahill1992}%
  \BibitemOpen
  \bibfield  {author} {\bibinfo {author} {\bibfnamefont {David~G.}\
  \bibnamefont {Cahill}}, \bibinfo {author} {\bibfnamefont {S.~K.}\
  \bibnamefont {Watson}}, \ and\ \bibinfo {author} {\bibfnamefont {R.~O.}\
  \bibnamefont {Pohl}},\ }\bibfield  {title} {Lower limit to the thermal
  conductivity of disordered crystals,\ }\href {\doibase
  10.1103/PhysRevB.46.6131} {\bibfield  {journal} {\bibinfo  {journal} {Phys.
  Rev. B}\ }\textbf {\bibinfo {volume} {46}},\ \bibinfo {pages} {6131--6140}
  (\bibinfo {year} {1992})}\BibitemShut {NoStop}%
\bibitem [{\citenamefont {Agne}\ \emph {et~al.}(2018)\citenamefont {Agne},
  \citenamefont {Hanus},\ and\ \citenamefont {Snyder}}]{Agne2018}%
  \BibitemOpen
  \bibfield  {author} {\bibinfo {author} {\bibfnamefont {Matthias~T.}\
  \bibnamefont {Agne}}, \bibinfo {author} {\bibfnamefont {Riley}\ \bibnamefont
  {Hanus}}, \ and\ \bibinfo {author} {\bibfnamefont {G.~Jeffrey}\ \bibnamefont
  {Snyder}},\ }\bibfield  {title} {Minimum thermal conductivity in the context
  of diffuson-mediated thermal transport,\ }\href {\doibase 10.1039/C7EE03256K}
  {\bibfield  {journal} {\bibinfo  {journal} {Energy Environ. Sci.}\ }\textbf
  {\bibinfo {volume} {11}},\ \bibinfo {pages} {609--616} (\bibinfo {year}
  {2018})}\BibitemShut {NoStop}%
\bibitem [{\citenamefont {Cahill}\ and\ \citenamefont
  {Pohl}(1988)}]{Cahill1988}%
  \BibitemOpen
  \bibfield  {author} {\bibinfo {author} {\bibfnamefont {D~G}\ \bibnamefont
  {Cahill}}\ and\ \bibinfo {author} {\bibfnamefont {R~O}\ \bibnamefont
  {Pohl}},\ }\bibfield  {title} {Lattice Vibrations and Heat Transport in
  Crystals and Glasses,\ }\href {\doibase 10.1146/annurev.pc.39.100188.000521}
  {\bibfield  {journal} {\bibinfo  {journal} {Annual Review of Physical
  Chemistry}\ }\textbf {\bibinfo {volume} {39}},\ \bibinfo {pages} {93--121}
  (\bibinfo {year} {1988})}\BibitemShut {NoStop}%
\bibitem [{\citenamefont {Cahill}\ and\ \citenamefont
  {Pohl}(1989)}]{Cahill1989}%
  \BibitemOpen
  \bibfield  {author} {\bibinfo {author} {\bibfnamefont {David~G.}\
  \bibnamefont {Cahill}}\ and\ \bibinfo {author} {\bibfnamefont {R.O.}\
  \bibnamefont {Pohl}},\ }\bibfield  {title} {Heat flow and lattice vibrations
  in glasses,\ }\href {\doibase https://doi.org/10.1016/0038-1098(89)90630-3}
  {\bibfield  {journal} {\bibinfo  {journal} {Solid State Communications}\
  }\textbf {\bibinfo {volume} {70}},\ \bibinfo {pages} {927 -- 930} (\bibinfo
  {year} {1989})}\BibitemShut {NoStop}%
\bibitem [{\citenamefont {Einstein}(1911)}]{Einstein1911}%
  \BibitemOpen
  \bibfield  {author} {\bibinfo {author} {\bibfnamefont {A.}~\bibnamefont
  {Einstein}},\ }\bibfield  {title} {Elementare Betrachtungen {\"u}ber die
  thermische Molekularbewegung in festen K{\"o}rpern,\ }\href {\doibase
  10.1002/andp.19113400903} {\bibfield  {journal} {\bibinfo  {journal} {Annalen
  der Physik}\ }\textbf {\bibinfo {volume} {340}},\ \bibinfo {pages} {679--694}
  (\bibinfo {year} {1911})}\BibitemShut {NoStop}%
\bibitem [{\citenamefont {Kittel}(2004)}]{kittel2004introduction}%
  \BibitemOpen
  \bibfield  {author} {\bibinfo {author} {\bibfnamefont {C.}~\bibnamefont
  {Kittel}},\ }\href {https://books.google.com/books?id=kym4QgAACAAJ} {\emph
  {\bibinfo {title} {Introduction to Solid State Physics}}}\ (\bibinfo
  {publisher} {Wiley},\ \bibinfo {year} {2004})\BibitemShut {NoStop}%
\bibitem [{\citenamefont {Allen}\ and\ \citenamefont
  {Feldman}(1989)}]{AllenPRL1989}%
  \BibitemOpen
  \bibfield  {author} {\bibinfo {author} {\bibfnamefont {Philip~B.}\
  \bibnamefont {Allen}}\ and\ \bibinfo {author} {\bibfnamefont {Joseph~L.}\
  \bibnamefont {Feldman}},\ }\bibfield  {title} {Thermal Conductivity of
  Glasses: Theory and Application to Amorphous {Si},\ }\href {\doibase
  10.1103/PhysRevLett.62.645} {\bibfield  {journal} {\bibinfo  {journal} {Phys.
  Rev. Lett.}\ }\textbf {\bibinfo {volume} {62}},\ \bibinfo {pages} {645--648}
  (\bibinfo {year} {1989})}\BibitemShut {NoStop}%
\bibitem [{\citenamefont {Allen}\ and\ \citenamefont
  {Feldman}(1993)}]{Allen1993disorder}%
  \BibitemOpen
  \bibfield  {author} {\bibinfo {author} {\bibfnamefont {Philip~B.}\
  \bibnamefont {Allen}}\ and\ \bibinfo {author} {\bibfnamefont {Joseph~L.}\
  \bibnamefont {Feldman}},\ }\bibfield  {title} {Thermal conductivity of
  disordered harmonic solids,\ }\href {\doibase 10.1103/PhysRevB.48.12581}
  {\bibfield  {journal} {\bibinfo  {journal} {Phys. Rev. B}\ }\textbf {\bibinfo
  {volume} {48}},\ \bibinfo {pages} {12581--12588} (\bibinfo {year}
  {1993})}\BibitemShut {NoStop}%
\bibitem [{\citenamefont {Allen}\ \emph {et~al.}(1999)\citenamefont {Allen},
  \citenamefont {Feldman}, \citenamefont {Fabian},\ and\ \citenamefont
  {Wooten}}]{allen1999diffusons}%
  \BibitemOpen
  \bibfield  {author} {\bibinfo {author} {\bibfnamefont {Philip~B}\
  \bibnamefont {Allen}}, \bibinfo {author} {\bibfnamefont {Joseph~L}\
  \bibnamefont {Feldman}}, \bibinfo {author} {\bibfnamefont {Jaroslav}\
  \bibnamefont {Fabian}}, \ and\ \bibinfo {author} {\bibfnamefont {Frederick}\
  \bibnamefont {Wooten}},\ }\bibfield  {title} {Diffusons, locons and
  propagons: Character of atomic vibrations in amorphous {Si},\ }\href@noop {}
  {\bibfield  {journal} {\bibinfo  {journal} {Philosophical Magazine B}\
  }\textbf {\bibinfo {volume} {79}},\ \bibinfo {pages} {1715--1731} (\bibinfo
  {year} {1999})}\BibitemShut {NoStop}%
\bibitem [{\citenamefont {Mukhopadhyay}\ \emph {et~al.}(2018)\citenamefont
  {Mukhopadhyay}, \citenamefont {Parker}, \citenamefont {Sales}, \citenamefont
  {Puretzky}, \citenamefont {McGuire},\ and\ \citenamefont
  {Lindsay}}]{Mukhopadhyay1455}%
  \BibitemOpen
  \bibfield  {author} {\bibinfo {author} {\bibfnamefont {Saikat}\ \bibnamefont
  {Mukhopadhyay}}, \bibinfo {author} {\bibfnamefont {David~S.}\ \bibnamefont
  {Parker}}, \bibinfo {author} {\bibfnamefont {Brian~C.}\ \bibnamefont
  {Sales}}, \bibinfo {author} {\bibfnamefont {Alexander~A.}\ \bibnamefont
  {Puretzky}}, \bibinfo {author} {\bibfnamefont {Michael~A.}\ \bibnamefont
  {McGuire}}, \ and\ \bibinfo {author} {\bibfnamefont {Lucas}\ \bibnamefont
  {Lindsay}},\ }\bibfield  {title} {Two-channel model for ultralow thermal
  conductivity of crystalline {Tl3VSe4},\ }\href {\doibase
  10.1126/science.aar8072} {\bibfield  {journal} {\bibinfo  {journal}
  {Science}\ }\textbf {\bibinfo {volume} {360}},\ \bibinfo {pages} {1455--1458}
  (\bibinfo {year} {2018})}\BibitemShut {NoStop}%
\bibitem [{\citenamefont {Luo}\ \emph {et~al.}(2020)\citenamefont {Luo},
  \citenamefont {Yang}, \citenamefont {Feng}, \citenamefont {Wang},\ and\
  \citenamefont {Ruan}}]{Luo:2020aa}%
  \BibitemOpen
  \bibfield  {author} {\bibinfo {author} {\bibfnamefont {Yixiu}\ \bibnamefont
  {Luo}}, \bibinfo {author} {\bibfnamefont {Xiaolong}\ \bibnamefont {Yang}},
  \bibinfo {author} {\bibfnamefont {Tianli}\ \bibnamefont {Feng}}, \bibinfo
  {author} {\bibfnamefont {Jingyang}\ \bibnamefont {Wang}}, \ and\ \bibinfo
  {author} {\bibfnamefont {Xiulin}\ \bibnamefont {Ruan}},\ }\bibfield  {title}
  {Vibrational hierarchy leads to dual-phonon transport in low thermal
  conductivity crystals,\ }\href {\doibase 10.1038/s41467-020-16371-w}
  {\bibfield  {journal} {\bibinfo  {journal} {Nature Communications}\ }\textbf
  {\bibinfo {volume} {11}},\ \bibinfo {pages} {2554} (\bibinfo {year}
  {2020})}\BibitemShut {NoStop}%
\bibitem [{\citenamefont {Xia}\ \emph {et~al.}(2020{\natexlab{a}})\citenamefont
  {Xia}, \citenamefont {Ozoli\ifmmode \mbox{\c{n}}\else
  \c{n}\fi{}\ifmmode~\check{s}\else \v{s}\fi{}},\ and\ \citenamefont
  {Wolverton}}]{PhysRevLett.125.085901}%
  \BibitemOpen
  \bibfield  {author} {\bibinfo {author} {\bibfnamefont {Yi}~\bibnamefont
  {Xia}}, \bibinfo {author} {\bibfnamefont {Vidvuds}\ \bibnamefont
  {Ozoli\ifmmode \mbox{\c{n}}\else \c{n}\fi{}\ifmmode~\check{s}\else
  \v{s}\fi{}}}, \ and\ \bibinfo {author} {\bibfnamefont {Chris}\ \bibnamefont
  {Wolverton}},\ }\bibfield  {title} {Microscopic Mechanisms of Glasslike
  Lattice Thermal Transport in Cubic {Cu}$_{12}${Sb}$_{4}${S}$_{13}$
  Tetrahedrites,\ }\href {\doibase 10.1103/PhysRevLett.125.085901} {\bibfield
  {journal} {\bibinfo  {journal} {Phys. Rev. Lett.}\ }\textbf {\bibinfo
  {volume} {125}},\ \bibinfo {pages} {085901} (\bibinfo {year}
  {2020}{\natexlab{a}})}\BibitemShut {NoStop}%
\bibitem [{\citenamefont {Hanus}\ \emph {et~al.}(2021)\citenamefont {Hanus},
  \citenamefont {George}, \citenamefont {Wood}, \citenamefont {Bonkowski},
  \citenamefont {Cheng}, \citenamefont {Abernathy}, \citenamefont {Manley},
  \citenamefont {Hautier}, \citenamefont {Snyder},\ and\ \citenamefont
  {Hermann}}]{Hanus2021}%
  \BibitemOpen
  \bibfield  {author} {\bibinfo {author} {\bibfnamefont {Riley}\ \bibnamefont
  {Hanus}}, \bibinfo {author} {\bibfnamefont {Janine}\ \bibnamefont {George}},
  \bibinfo {author} {\bibfnamefont {Max}\ \bibnamefont {Wood}}, \bibinfo
  {author} {\bibfnamefont {Alexander}\ \bibnamefont {Bonkowski}}, \bibinfo
  {author} {\bibfnamefont {Yongqiang}\ \bibnamefont {Cheng}}, \bibinfo {author}
  {\bibfnamefont {Douglas~L.}\ \bibnamefont {Abernathy}}, \bibinfo {author}
  {\bibfnamefont {Michael~E.}\ \bibnamefont {Manley}}, \bibinfo {author}
  {\bibfnamefont {Geoffroy}\ \bibnamefont {Hautier}}, \bibinfo {author}
  {\bibfnamefont {G.~Jeffrey}\ \bibnamefont {Snyder}}, \ and\ \bibinfo {author}
  {\bibfnamefont {Rapha{\"e}l~P.}\ \bibnamefont {Hermann}},\ }\bibfield
  {title} {Uncovering design principles for amorphous-like heat conduction
  using two-channel lattice dynamics,\ }\href {\doibase
  https://doi.org/10.1016/j.mtphys.2021.100344} {\bibfield  {journal} {\bibinfo
   {journal} {Materials Today Physics}\ }\textbf {\bibinfo {volume} {18}},\
  \bibinfo {pages} {100344} (\bibinfo {year} {2021})}\BibitemShut {NoStop}%
\bibitem [{\citenamefont {Kurosaki}\ \emph {et~al.}(2005)\citenamefont
  {Kurosaki}, \citenamefont {Kosuga}, \citenamefont {Goto}, \citenamefont
  {Muta},\ and\ \citenamefont {Yamanaka}}]{AgTlTe2005}%
  \BibitemOpen
  \bibfield  {author} {\bibinfo {author} {\bibfnamefont {Ken}\ \bibnamefont
  {Kurosaki}}, \bibinfo {author} {\bibfnamefont {Atsuko}\ \bibnamefont
  {Kosuga}}, \bibinfo {author} {\bibfnamefont {Keita}\ \bibnamefont {Goto}},
  \bibinfo {author} {\bibfnamefont {Hiroaki}\ \bibnamefont {Muta}}, \ and\
  \bibinfo {author} {\bibfnamefont {Shinsuke}\ \bibnamefont {Yamanaka}},\
  }\bibfield  {title} {{Ag}$_9${TlTe}$_5$ and {AgTlTe}: high {ZT} materials
  with extremely low thermal conductivity,\ }in\ \href@noop {} {\emph {\bibinfo
  {booktitle} {ICT 2005. 24th International Conference on Thermoelectrics,
  2005.}}}\ (\bibinfo {organization} {IEEE},\ \bibinfo {year} {2005})\ pp.\
  \bibinfo {pages} {323--326}\BibitemShut {NoStop}%
\bibitem [{\citenamefont {Liu}\ \emph {et~al.}(2012)\citenamefont {Liu},
  \citenamefont {Shi}, \citenamefont {Xu}, \citenamefont {Zhang}, \citenamefont
  {Zhang}, \citenamefont {Chen}, \citenamefont {Li}, \citenamefont {Uher},
  \citenamefont {Day},\ and\ \citenamefont {Snyder}}]{Liu:2012aa}%
  \BibitemOpen
  \bibfield  {author} {\bibinfo {author} {\bibfnamefont {Huili}\ \bibnamefont
  {Liu}}, \bibinfo {author} {\bibfnamefont {Xun}\ \bibnamefont {Shi}}, \bibinfo
  {author} {\bibfnamefont {Fangfang}\ \bibnamefont {Xu}}, \bibinfo {author}
  {\bibfnamefont {Linlin}\ \bibnamefont {Zhang}}, \bibinfo {author}
  {\bibfnamefont {Wenqing}\ \bibnamefont {Zhang}}, \bibinfo {author}
  {\bibfnamefont {Lidong}\ \bibnamefont {Chen}}, \bibinfo {author}
  {\bibfnamefont {Qiang}\ \bibnamefont {Li}}, \bibinfo {author} {\bibfnamefont
  {Ctirad}\ \bibnamefont {Uher}}, \bibinfo {author} {\bibfnamefont {Tristan}\
  \bibnamefont {Day}}, \ and\ \bibinfo {author} {\bibfnamefont {G.~Jeffrey}\
  \bibnamefont {Snyder}},\ }\bibfield  {title} {Copper ion liquid-like
  thermoelectrics,\ }\href {\doibase 10.1038/nmat3273} {\bibfield  {journal}
  {\bibinfo  {journal} {Nature Materials}\ }\textbf {\bibinfo {volume} {11}},\
  \bibinfo {pages} {422--425} (\bibinfo {year} {2012})}\BibitemShut {NoStop}%
\bibitem [{\citenamefont {Lu}\ \emph {et~al.}(2013)\citenamefont {Lu},
  \citenamefont {Morelli}, \citenamefont {Xia}, \citenamefont {Zhou},
  \citenamefont {Ozolins}, \citenamefont {Chi}, \citenamefont {Zhou},\ and\
  \citenamefont {Uher}}]{Lu2013}%
  \BibitemOpen
  \bibfield  {author} {\bibinfo {author} {\bibfnamefont {Xu}~\bibnamefont
  {Lu}}, \bibinfo {author} {\bibfnamefont {Donald~T.}\ \bibnamefont {Morelli}},
  \bibinfo {author} {\bibfnamefont {Yi}~\bibnamefont {Xia}}, \bibinfo {author}
  {\bibfnamefont {Fei}\ \bibnamefont {Zhou}}, \bibinfo {author} {\bibfnamefont
  {Vidvuds}\ \bibnamefont {Ozolins}}, \bibinfo {author} {\bibfnamefont {Hang}\
  \bibnamefont {Chi}}, \bibinfo {author} {\bibfnamefont {Xiaoyuan}\
  \bibnamefont {Zhou}}, \ and\ \bibinfo {author} {\bibfnamefont {Ctirad}\
  \bibnamefont {Uher}},\ }\bibfield  {title} {High Performance
  Thermoelectricity in Earth-Abundant Compounds Based on Natural Mineral
  Tetrahedrites,\ }\href {\doibase 10.1002/aenm.201200650} {\bibfield
  {journal} {\bibinfo  {journal} {Advanced Energy Materials}\ }\textbf
  {\bibinfo {volume} {3}},\ \bibinfo {pages} {342--348} (\bibinfo {year}
  {2013})}\BibitemShut {NoStop}%
\bibitem [{\citenamefont {Lin}\ \emph {et~al.}(2016)\citenamefont {Lin},
  \citenamefont {Tan}, \citenamefont {Shen}, \citenamefont {Hao}, \citenamefont
  {Wu}, \citenamefont {Calta}, \citenamefont {Malliakas}, \citenamefont {Wang},
  \citenamefont {Uher}, \citenamefont {Wolverton},\ and\ \citenamefont
  {Kanatzidis}}]{CsAg5Te32016}%
  \BibitemOpen
  \bibfield  {author} {\bibinfo {author} {\bibfnamefont {Hua}\ \bibnamefont
  {Lin}}, \bibinfo {author} {\bibfnamefont {Gangjian}\ \bibnamefont {Tan}},
  \bibinfo {author} {\bibfnamefont {Jin-Ni}\ \bibnamefont {Shen}}, \bibinfo
  {author} {\bibfnamefont {Shiqiang}\ \bibnamefont {Hao}}, \bibinfo {author}
  {\bibfnamefont {Li-Ming}\ \bibnamefont {Wu}}, \bibinfo {author}
  {\bibfnamefont {Nicholas}\ \bibnamefont {Calta}}, \bibinfo {author}
  {\bibfnamefont {Christos}\ \bibnamefont {Malliakas}}, \bibinfo {author}
  {\bibfnamefont {Si}~\bibnamefont {Wang}}, \bibinfo {author} {\bibfnamefont
  {Ctirad}\ \bibnamefont {Uher}}, \bibinfo {author} {\bibfnamefont
  {Christopher}\ \bibnamefont {Wolverton}}, \ and\ \bibinfo {author}
  {\bibfnamefont {Mercouri~G.}\ \bibnamefont {Kanatzidis}},\ }\bibfield
  {title} {Concerted Rattling in {CsAg5Te3} Leading to Ultralow Thermal
  Conductivity and High Thermoelectric Performance,\ }\href {\doibase
  https://doi.org/10.1002/anie.201605015} {\bibfield  {journal} {\bibinfo
  {journal} {Angewandte Chemie International Edition}\ }\textbf {\bibinfo
  {volume} {55}},\ \bibinfo {pages} {11431--11436} (\bibinfo {year}
  {2016})}\BibitemShut {NoStop}%
\bibitem [{\citenamefont {Lin}\ \emph {et~al.}(2017)\citenamefont {Lin},
  \citenamefont {Li}, \citenamefont {Li}, \citenamefont {Zhang}, \citenamefont
  {Chen}, \citenamefont {Xu}, \citenamefont {Chen},\ and\ \citenamefont
  {Pei}}]{LIN2017816}%
  \BibitemOpen
  \bibfield  {author} {\bibinfo {author} {\bibfnamefont {Siqi}\ \bibnamefont
  {Lin}}, \bibinfo {author} {\bibfnamefont {Wen}\ \bibnamefont {Li}}, \bibinfo
  {author} {\bibfnamefont {Shasha}\ \bibnamefont {Li}}, \bibinfo {author}
  {\bibfnamefont {Xinyue}\ \bibnamefont {Zhang}}, \bibinfo {author}
  {\bibfnamefont {Zhiwei}\ \bibnamefont {Chen}}, \bibinfo {author}
  {\bibfnamefont {Yidong}\ \bibnamefont {Xu}}, \bibinfo {author} {\bibfnamefont
  {Yue}\ \bibnamefont {Chen}}, \ and\ \bibinfo {author} {\bibfnamefont
  {Yanzhong}\ \bibnamefont {Pei}},\ }\bibfield  {title} {High Thermoelectric
  Performance of {Ag9GaSe6} Enabled by Low Cutoff Frequency of Acoustic
  Phonons,\ }\href {\doibase https://doi.org/10.1016/j.joule.2017.09.006}
  {\bibfield  {journal} {\bibinfo  {journal} {Joule}\ }\textbf {\bibinfo
  {volume} {1}},\ \bibinfo {pages} {816--830} (\bibinfo {year}
  {2017})}\BibitemShut {NoStop}%
\bibitem [{\citenamefont {Slack}(1973)}]{slack1973nonmetallic}%
  \BibitemOpen
  \bibfield  {author} {\bibinfo {author} {\bibfnamefont {Glen~A}\ \bibnamefont
  {Slack}},\ }\bibfield  {title} {Nonmetallic crystals with high thermal
  conductivity,\ }\href@noop {} {\bibfield  {journal} {\bibinfo  {journal}
  {Journal of Physics and Chemistry of Solids}\ }\textbf {\bibinfo {volume}
  {34}},\ \bibinfo {pages} {321--335} (\bibinfo {year} {1973})}\BibitemShut
  {NoStop}%
\bibitem [{\citenamefont {Simoncelli}\ \emph {et~al.}(2019)\citenamefont
  {Simoncelli}, \citenamefont {Marzari},\ and\ \citenamefont
  {Mauri}}]{Simoncelli2019}%
  \BibitemOpen
  \bibfield  {author} {\bibinfo {author} {\bibfnamefont {Michele}\ \bibnamefont
  {Simoncelli}}, \bibinfo {author} {\bibfnamefont {Nicola}\ \bibnamefont
  {Marzari}}, \ and\ \bibinfo {author} {\bibfnamefont {Francesco}\ \bibnamefont
  {Mauri}},\ }\bibfield  {title} {Unified theory of thermal transport in
  crystals and glasses,\ }\href {\doibase 10.1038/s41567-019-0520-x} {\bibfield
   {journal} {\bibinfo  {journal} {Nature Physics}\ }\textbf {\bibinfo {volume}
  {15}},\ \bibinfo {pages} {809--813} (\bibinfo {year} {2019})}\BibitemShut
  {NoStop}%
\bibitem [{\citenamefont {Isaeva}\ \emph {et~al.}(2019)\citenamefont {Isaeva},
  \citenamefont {Barbalinardo}, \citenamefont {Donadio},\ and\ \citenamefont
  {Baroni}}]{Isaeva2019}%
  \BibitemOpen
  \bibfield  {author} {\bibinfo {author} {\bibfnamefont {Leyla}\ \bibnamefont
  {Isaeva}}, \bibinfo {author} {\bibfnamefont {Giuseppe}\ \bibnamefont
  {Barbalinardo}}, \bibinfo {author} {\bibfnamefont {Davide}\ \bibnamefont
  {Donadio}}, \ and\ \bibinfo {author} {\bibfnamefont {Stefano}\ \bibnamefont
  {Baroni}},\ }\bibfield  {title} {Modeling heat transport in crystals and
  glasses from a unified lattice-dynamical approach,\ }\href {\doibase
  10.1038/s41467-019-11572-4} {\bibfield  {journal} {\bibinfo  {journal}
  {Nature Communications}\ }\textbf {\bibinfo {volume} {10}},\ \bibinfo {pages}
  {3853} (\bibinfo {year} {2019})}\BibitemShut {NoStop}%
\bibitem [{\citenamefont {Simoncelli}\ \emph {et~al.}(2022)\citenamefont
  {Simoncelli}, \citenamefont {Marzari},\ and\ \citenamefont
  {Mauri}}]{wigner2022}%
  \BibitemOpen
  \bibfield  {author} {\bibinfo {author} {\bibfnamefont {Michele}\ \bibnamefont
  {Simoncelli}}, \bibinfo {author} {\bibfnamefont {Nicola}\ \bibnamefont
  {Marzari}}, \ and\ \bibinfo {author} {\bibfnamefont {Francesco}\ \bibnamefont
  {Mauri}},\ }\bibfield  {title} {Wigner Formulation of Thermal Transport in
  Solids,\ }\href {\doibase 10.1103/PhysRevX.12.041011} {\bibfield  {journal}
  {\bibinfo  {journal} {Phys. Rev. X}\ }\textbf {\bibinfo {volume} {12}},\
  \bibinfo {pages} {041011} (\bibinfo {year} {2022})}\BibitemShut {NoStop}%
\bibitem [{\citenamefont {Togo}(2018)}]{phonondbtogo}%
  \BibitemOpen
  \bibfield  {author} {\bibinfo {author} {\bibfnamefont {Atsushi}\ \bibnamefont
  {Togo}},\ }\href@noop {} {Phonon database at Kyoto university},\ \bibinfo
  {howpublished} {http://phonondb.mtl.kyoto-u.ac.jp/} (\bibinfo {year}
  {2018})\BibitemShut {NoStop}%
\bibitem [{\citenamefont {Togo}\ and\ \citenamefont {Tanaka}(2015)}]{Togo2015}%
  \BibitemOpen
  \bibfield  {author} {\bibinfo {author} {\bibfnamefont {Atsushi}\ \bibnamefont
  {Togo}}\ and\ \bibinfo {author} {\bibfnamefont {Isao}\ \bibnamefont
  {Tanaka}},\ }\bibfield  {title} {First principles phonon calculations in
  materials science,\ }\href {\doibase
  http://doi.org/10.1016/j.scriptamat.2015.07.021} {\bibfield  {journal}
  {\bibinfo  {journal} {Scripta Materialia}\ }\textbf {\bibinfo {volume}
  {108}},\ \bibinfo {pages} {1 -- 5} (\bibinfo {year} {2015})}\BibitemShut
  {NoStop}%
\bibitem [{\citenamefont {Jain}\ \emph {et~al.}(2013)\citenamefont {Jain},
  \citenamefont {Ong}, \citenamefont {Hautier}, \citenamefont {Chen},
  \citenamefont {Richards}, \citenamefont {Dacek}, \citenamefont {Cholia},
  \citenamefont {Gunter}, \citenamefont {Skinner}, \citenamefont {Ceder},\ and\
  \citenamefont {Persson}}]{matproj}%
  \BibitemOpen
  \bibfield  {author} {\bibinfo {author} {\bibfnamefont {Anubhav}\ \bibnamefont
  {Jain}}, \bibinfo {author} {\bibfnamefont {Shyue~Ping}\ \bibnamefont {Ong}},
  \bibinfo {author} {\bibfnamefont {Geoffroy}\ \bibnamefont {Hautier}},
  \bibinfo {author} {\bibfnamefont {Wei}\ \bibnamefont {Chen}}, \bibinfo
  {author} {\bibfnamefont {William~Davidson}\ \bibnamefont {Richards}},
  \bibinfo {author} {\bibfnamefont {Stephen}\ \bibnamefont {Dacek}}, \bibinfo
  {author} {\bibfnamefont {Shreyas}\ \bibnamefont {Cholia}}, \bibinfo {author}
  {\bibfnamefont {Dan}\ \bibnamefont {Gunter}}, \bibinfo {author}
  {\bibfnamefont {David}\ \bibnamefont {Skinner}}, \bibinfo {author}
  {\bibfnamefont {Gerbrand}\ \bibnamefont {Ceder}}, \ and\ \bibinfo {author}
  {\bibfnamefont {Kristin~A.}\ \bibnamefont {Persson}},\ }\bibfield  {title}
  {{Commentary: The Materials Project}: A materials genome approach to
  accelerating materials innovation,\ }\href {\doibase 10.1063/1.4812323}
  {\bibfield  {journal} {\bibinfo  {journal} {APL Materials}\ }\textbf
  {\bibinfo {volume} {1}},\ \bibinfo {pages} {011002} (\bibinfo {year}
  {2013})},\ \Eprint {http://arxiv.org/abs/https://doi.org/10.1063/1.4812323}
  {https://doi.org/10.1063/1.4812323} \BibitemShut {NoStop}%
\bibitem [{\citenamefont {Ong}\ \emph {et~al.}(2013)\citenamefont {Ong},
  \citenamefont {Richards}, \citenamefont {Jain}, \citenamefont {Hautier},
  \citenamefont {Kocher}, \citenamefont {Cholia}, \citenamefont {Gunter},
  \citenamefont {Chevrier}, \citenamefont {Persson},\ and\ \citenamefont
  {Ceder}}]{pymatgen}%
  \BibitemOpen
  \bibfield  {author} {\bibinfo {author} {\bibfnamefont {Shyue~Ping}\
  \bibnamefont {Ong}}, \bibinfo {author} {\bibfnamefont {William~Davidson}\
  \bibnamefont {Richards}}, \bibinfo {author} {\bibfnamefont {Anubhav}\
  \bibnamefont {Jain}}, \bibinfo {author} {\bibfnamefont {Geoffroy}\
  \bibnamefont {Hautier}}, \bibinfo {author} {\bibfnamefont {Michael}\
  \bibnamefont {Kocher}}, \bibinfo {author} {\bibfnamefont {Shreyas}\
  \bibnamefont {Cholia}}, \bibinfo {author} {\bibfnamefont {Dan}\ \bibnamefont
  {Gunter}}, \bibinfo {author} {\bibfnamefont {Vincent~L.}\ \bibnamefont
  {Chevrier}}, \bibinfo {author} {\bibfnamefont {Kristin~A.}\ \bibnamefont
  {Persson}}, \ and\ \bibinfo {author} {\bibfnamefont {Gerbrand}\ \bibnamefont
  {Ceder}},\ }\bibfield  {title} {{Python Materials Genomics (pymatgen)}: A
  robust, open-source python library for materials analysis,\ }\href {\doibase
  https://doi.org/10.1016/j.commatsci.2012.10.028} {\bibfield  {journal}
  {\bibinfo  {journal} {Computational Materials Science}\ }\textbf {\bibinfo
  {volume} {68}},\ \bibinfo {pages} {314--319} (\bibinfo {year}
  {2013})}\BibitemShut {NoStop}%
\bibitem [{\citenamefont {Ong}\ \emph {et~al.}(2015)\citenamefont {Ong},
  \citenamefont {Cholia}, \citenamefont {Jain}, \citenamefont {Brafman},
  \citenamefont {Gunter}, \citenamefont {Ceder},\ and\ \citenamefont
  {Persson}}]{mpapi}%
  \BibitemOpen
  \bibfield  {author} {\bibinfo {author} {\bibfnamefont {Shyue~Ping}\
  \bibnamefont {Ong}}, \bibinfo {author} {\bibfnamefont {Shreyas}\ \bibnamefont
  {Cholia}}, \bibinfo {author} {\bibfnamefont {Anubhav}\ \bibnamefont {Jain}},
  \bibinfo {author} {\bibfnamefont {Miriam}\ \bibnamefont {Brafman}}, \bibinfo
  {author} {\bibfnamefont {Dan}\ \bibnamefont {Gunter}}, \bibinfo {author}
  {\bibfnamefont {Gerbrand}\ \bibnamefont {Ceder}}, \ and\ \bibinfo {author}
  {\bibfnamefont {Kristin~A.}\ \bibnamefont {Persson}},\ }\bibfield  {title}
  {{The Materials Application Programming Interface (API): A simple, flexible
  and efficient API for materials data based on REpresentational State Transfer
  (REST) principles},\ }\href {\doibase
  https://doi.org/10.1016/j.commatsci.2014.10.037} {\bibfield  {journal}
  {\bibinfo  {journal} {Computational Materials Science}\ }\textbf {\bibinfo
  {volume} {97}},\ \bibinfo {pages} {209--215} (\bibinfo {year}
  {2015})}\BibitemShut {NoStop}%
\bibitem [{\citenamefont {Slack}(1979{\natexlab{b}})}]{slack1979thermal}%
  \BibitemOpen
  \bibfield  {author} {\bibinfo {author} {\bibfnamefont {Glen~A}\ \bibnamefont
  {Slack}},\ }\bibfield  {title} {The thermal conductivity of nonmetallic
  crystals,\ }\href@noop {} {\bibfield  {journal} {\bibinfo  {journal} {Solid
  state physics}\ }\textbf {\bibinfo {volume} {34}},\ \bibinfo {pages} {1--71}
  (\bibinfo {year} {1979}{\natexlab{b}})}\BibitemShut {NoStop}%
\bibitem [{\citenamefont {Toberer}\ \emph {et~al.}(2011)\citenamefont
  {Toberer}, \citenamefont {Zevalkink},\ and\ \citenamefont
  {Snyder}}]{toberer}%
  \BibitemOpen
  \bibfield  {author} {\bibinfo {author} {\bibfnamefont {Eric~S.}\ \bibnamefont
  {Toberer}}, \bibinfo {author} {\bibfnamefont {Alex}\ \bibnamefont
  {Zevalkink}}, \ and\ \bibinfo {author} {\bibfnamefont {G.~Jeffrey}\
  \bibnamefont {Snyder}},\ }\bibfield  {title} {Phonon engineering through
  crystal chemistry,\ }\href {\doibase 10.1039/C1JM11754H} {\bibfield
  {journal} {\bibinfo  {journal} {J. Mater. Chem.}\ }\textbf {\bibinfo {volume}
  {21}},\ \bibinfo {pages} {15843--15852} (\bibinfo {year} {2011})}\BibitemShut
  {NoStop}%
\bibitem [{\citenamefont {Chen}\ \emph {et~al.}(2020)\citenamefont {Chen},
  \citenamefont {Song}, \citenamefont {Ravichandran}, \citenamefont {Zheng},
  \citenamefont {Chen}, \citenamefont {Lee}, \citenamefont {Sun}, \citenamefont
  {Li}, \citenamefont {Gamage}, \citenamefont {Tian}, \citenamefont {Ding},
  \citenamefont {Song}, \citenamefont {Rai}, \citenamefont {Wu}, \citenamefont
  {Koirala}, \citenamefont {Schmidt}, \citenamefont {Watanabe}, \citenamefont
  {Lv}, \citenamefont {Ren}, \citenamefont {Shi}, \citenamefont {Cahill},
  \citenamefont {Taniguchi}, \citenamefont {Broido},\ and\ \citenamefont
  {Chen}}]{kechenBN}%
  \BibitemOpen
  \bibfield  {author} {\bibinfo {author} {\bibfnamefont {Ke}~\bibnamefont
  {Chen}}, \bibinfo {author} {\bibfnamefont {Bai}\ \bibnamefont {Song}},
  \bibinfo {author} {\bibfnamefont {Navaneetha~K.}\ \bibnamefont
  {Ravichandran}}, \bibinfo {author} {\bibfnamefont {Qiye}\ \bibnamefont
  {Zheng}}, \bibinfo {author} {\bibfnamefont {Xi}~\bibnamefont {Chen}},
  \bibinfo {author} {\bibfnamefont {Hwijong}\ \bibnamefont {Lee}}, \bibinfo
  {author} {\bibfnamefont {Haoran}\ \bibnamefont {Sun}}, \bibinfo {author}
  {\bibfnamefont {Sheng}\ \bibnamefont {Li}}, \bibinfo {author} {\bibfnamefont
  {Geethal Amila Gamage~Udalamatta}\ \bibnamefont {Gamage}}, \bibinfo {author}
  {\bibfnamefont {Fei}\ \bibnamefont {Tian}}, \bibinfo {author} {\bibfnamefont
  {Zhiwei}\ \bibnamefont {Ding}}, \bibinfo {author} {\bibfnamefont {Qichen}\
  \bibnamefont {Song}}, \bibinfo {author} {\bibfnamefont {Akash}\ \bibnamefont
  {Rai}}, \bibinfo {author} {\bibfnamefont {Hanlin}\ \bibnamefont {Wu}},
  \bibinfo {author} {\bibfnamefont {Pawan}\ \bibnamefont {Koirala}}, \bibinfo
  {author} {\bibfnamefont {Aaron~J.}\ \bibnamefont {Schmidt}}, \bibinfo
  {author} {\bibfnamefont {Kenji}\ \bibnamefont {Watanabe}}, \bibinfo {author}
  {\bibfnamefont {Bing}\ \bibnamefont {Lv}}, \bibinfo {author} {\bibfnamefont
  {Zhifeng}\ \bibnamefont {Ren}}, \bibinfo {author} {\bibfnamefont
  {Li}~\bibnamefont {Shi}}, \bibinfo {author} {\bibfnamefont {David~G.}\
  \bibnamefont {Cahill}}, \bibinfo {author} {\bibfnamefont {Takashi}\
  \bibnamefont {Taniguchi}}, \bibinfo {author} {\bibfnamefont {David}\
  \bibnamefont {Broido}}, \ and\ \bibinfo {author} {\bibfnamefont {Gang}\
  \bibnamefont {Chen}},\ }\bibfield  {title} {Ultrahigh thermal conductivity in
  isotope-enriched cubic boron nitride,\ }\href {\doibase
  10.1126/science.aaz6149} {\bibfield  {journal} {\bibinfo  {journal}
  {Science}\ }\textbf {\bibinfo {volume} {367}},\ \bibinfo {pages} {555--559}
  (\bibinfo {year} {2020})},\ \Eprint
  {http://arxiv.org/abs/https://www.science.org/doi/pdf/10.1126/science.aaz6149}
  {https://www.science.org/doi/pdf/10.1126/science.aaz6149} \BibitemShut
  {NoStop}%
\bibitem [{\citenamefont {Hardy}(1963)}]{Hardy1963}%
  \BibitemOpen
  \bibfield  {author} {\bibinfo {author} {\bibfnamefont {Robert~J.}\
  \bibnamefont {Hardy}},\ }\bibfield  {title} {Energy-Flux Operator for a
  Lattice,\ }\href {\doibase 10.1103/PhysRev.132.168} {\bibfield  {journal}
  {\bibinfo  {journal} {Phys. Rev.}\ }\textbf {\bibinfo {volume} {132}},\
  \bibinfo {pages} {168--177} (\bibinfo {year} {1963})}\BibitemShut {NoStop}%
\bibitem [{\citenamefont {Wu}\ \emph {et~al.}(2019)\citenamefont {Wu},
  \citenamefont {Enamullah},\ and\ \citenamefont {Huang}}]{Minghui2019}%
  \BibitemOpen
  \bibfield  {author} {\bibinfo {author} {\bibfnamefont {Minghui}\ \bibnamefont
  {Wu}}, \bibinfo {author} {\bibnamefont {Enamullah}}, \ and\ \bibinfo {author}
  {\bibfnamefont {Li}~\bibnamefont {Huang}},\ }\bibfield  {title} {Unusual
  lattice thermal conductivity in the simple crystalline compounds
  {TlXTe}$_{2}${(X=Ga,In}),\ }\href {\doibase 10.1103/PhysRevB.100.075207}
  {\bibfield  {journal} {\bibinfo  {journal} {Phys. Rev. B}\ }\textbf {\bibinfo
  {volume} {100}},\ \bibinfo {pages} {075207} (\bibinfo {year}
  {2019})}\BibitemShut {NoStop}%
\bibitem [{\citenamefont {Jana}\ \emph {et~al.}(2017)\citenamefont {Jana},
  \citenamefont {Pal}, \citenamefont {Warankar}, \citenamefont {Mandal},
  \citenamefont {Waghmare},\ and\ \citenamefont {Biswas}}]{TlInTe2}%
  \BibitemOpen
  \bibfield  {author} {\bibinfo {author} {\bibfnamefont {Manoj~K.}\
  \bibnamefont {Jana}}, \bibinfo {author} {\bibfnamefont {Koushik}\
  \bibnamefont {Pal}}, \bibinfo {author} {\bibfnamefont {Avinash}\ \bibnamefont
  {Warankar}}, \bibinfo {author} {\bibfnamefont {Pankaj}\ \bibnamefont
  {Mandal}}, \bibinfo {author} {\bibfnamefont {Umesh~V.}\ \bibnamefont
  {Waghmare}}, \ and\ \bibinfo {author} {\bibfnamefont {Kanishka}\ \bibnamefont
  {Biswas}},\ }\bibfield  {title} {Intrinsic Rattler-Induced Low Thermal
  Conductivity in Zintl Type {TlInTe2},\ }\bibfield  {booktitle} {\emph
  {\bibinfo {booktitle} {Journal of the American Chemical Society}},\ }\href
  {\doibase 10.1021/jacs.7b01434} {\bibfield  {journal} {\bibinfo  {journal}
  {Journal of the American Chemical Society}\ }\textbf {\bibinfo {volume}
  {139}},\ \bibinfo {pages} {4350--4353} (\bibinfo {year} {2017})}\BibitemShut
  {NoStop}%
\bibitem [{\citenamefont {Pei}\ \emph {et~al.}(2014)\citenamefont {Pei},
  \citenamefont {Wu}, \citenamefont {Wu}, \citenamefont {Zheng},\ and\
  \citenamefont {He}}]{BiCuSeO}%
  \BibitemOpen
  \bibfield  {author} {\bibinfo {author} {\bibfnamefont {Yan-Ling}\
  \bibnamefont {Pei}}, \bibinfo {author} {\bibfnamefont {Haijun}\ \bibnamefont
  {Wu}}, \bibinfo {author} {\bibfnamefont {Di}~\bibnamefont {Wu}}, \bibinfo
  {author} {\bibfnamefont {Fengshan}\ \bibnamefont {Zheng}}, \ and\ \bibinfo
  {author} {\bibfnamefont {Jiaqing}\ \bibnamefont {He}},\ }\bibfield  {title}
  {High Thermoelectric Performance Realized in a {BiCuSeO} System by Improving
  Carrier Mobility through {3D} Modulation Doping,\ }\bibfield  {booktitle}
  {\emph {\bibinfo {booktitle} {Journal of the American Chemical Society}},\
  }\href {\doibase 10.1021/ja507945h} {\bibfield  {journal} {\bibinfo
  {journal} {Journal of the American Chemical Society}\ }\textbf {\bibinfo
  {volume} {136}},\ \bibinfo {pages} {13902--13908} (\bibinfo {year}
  {2014})}\BibitemShut {NoStop}%
\bibitem [{\citenamefont {Tyagi}\ \emph {et~al.}(2014)\citenamefont {Tyagi},
  \citenamefont {Gahtori}, \citenamefont {Bathula}, \citenamefont {Srivastava},
  \citenamefont {Shukla}, \citenamefont {Auluck},\ and\ \citenamefont
  {Dhar}}]{Cu3SbSe3}%
  \BibitemOpen
  \bibfield  {author} {\bibinfo {author} {\bibfnamefont {Kriti}\ \bibnamefont
  {Tyagi}}, \bibinfo {author} {\bibfnamefont {Bhasker}\ \bibnamefont
  {Gahtori}}, \bibinfo {author} {\bibfnamefont {Sivaiah}\ \bibnamefont
  {Bathula}}, \bibinfo {author} {\bibfnamefont {A.~K.}\ \bibnamefont
  {Srivastava}}, \bibinfo {author} {\bibfnamefont {A.~K.}\ \bibnamefont
  {Shukla}}, \bibinfo {author} {\bibfnamefont {Sushil}\ \bibnamefont {Auluck}},
  \ and\ \bibinfo {author} {\bibfnamefont {Ajay}\ \bibnamefont {Dhar}},\
  }\bibfield  {title} {Thermoelectric properties of {Cu3SbSe3} with
  intrinsically ultralow lattice thermal conductivity,\ }\href {\doibase
  10.1039/C4TA02590C} {\bibfield  {journal} {\bibinfo  {journal} {J. Mater.
  Chem. A}\ }\textbf {\bibinfo {volume} {2}},\ \bibinfo {pages} {15829--15835}
  (\bibinfo {year} {2014})}\BibitemShut {NoStop}%
\bibitem [{\citenamefont {Nolas}\ \emph {et~al.}(1998)\citenamefont {Nolas},
  \citenamefont {Cohn}, \citenamefont {Slack},\ and\ \citenamefont
  {Schujman}}]{SGGNolas}%
  \BibitemOpen
  \bibfield  {author} {\bibinfo {author} {\bibfnamefont {G.~S.}\ \bibnamefont
  {Nolas}}, \bibinfo {author} {\bibfnamefont {J.~L.}\ \bibnamefont {Cohn}},
  \bibinfo {author} {\bibfnamefont {G.~A.}\ \bibnamefont {Slack}}, \ and\
  \bibinfo {author} {\bibfnamefont {S.~B.}\ \bibnamefont {Schujman}},\
  }\bibfield  {title} {Semiconducting {Ge} clathrates: Promising candidates for
  thermoelectric applications,\ }\href {\doibase 10.1063/1.121747} {\bibfield
  {journal} {\bibinfo  {journal} {Applied Physics Letters}\ }\textbf {\bibinfo
  {volume} {73}},\ \bibinfo {pages} {178--180} (\bibinfo {year} {1998})},\
  \Eprint {http://arxiv.org/abs/https://doi.org/10.1063/1.121747}
  {https://doi.org/10.1063/1.121747} \BibitemShut {NoStop}%
\bibitem [{\citenamefont {Belsky}\ \emph {et~al.}(2002)\citenamefont {Belsky},
  \citenamefont {Hellenbrandt}, \citenamefont {Karen},\ and\ \citenamefont
  {Luksch}}]{icsd}%
  \BibitemOpen
  \bibfield  {author} {\bibinfo {author} {\bibfnamefont {Alec}\ \bibnamefont
  {Belsky}}, \bibinfo {author} {\bibfnamefont {Mariette}\ \bibnamefont
  {Hellenbrandt}}, \bibinfo {author} {\bibfnamefont {Vicky~Lynn}\ \bibnamefont
  {Karen}}, \ and\ \bibinfo {author} {\bibfnamefont {Peter}\ \bibnamefont
  {Luksch}},\ }\bibfield  {title} {New developments in the {Inorganic Crystal
  Structure Database (ICSD)}: accessibility in support of materials research
  and design,\ }\href@noop {} {\bibfield  {journal} {\bibinfo  {journal} {Acta
  Crystallographica Section B: Structural Science}\ }\textbf {\bibinfo {volume}
  {58}},\ \bibinfo {pages} {364--369} (\bibinfo {year} {2002})}\BibitemShut
  {NoStop}%
\bibitem [{\citenamefont {Chen}\ \emph {et~al.}(2019)\citenamefont {Chen},
  \citenamefont {Ye}, \citenamefont {Zuo}, \citenamefont {Zheng},\ and\
  \citenamefont {Ong}}]{Chen2019}%
  \BibitemOpen
  \bibfield  {author} {\bibinfo {author} {\bibfnamefont {Chi}\ \bibnamefont
  {Chen}}, \bibinfo {author} {\bibfnamefont {Weike}\ \bibnamefont {Ye}},
  \bibinfo {author} {\bibfnamefont {Yunxing}\ \bibnamefont {Zuo}}, \bibinfo
  {author} {\bibfnamefont {Chen}\ \bibnamefont {Zheng}}, \ and\ \bibinfo
  {author} {\bibfnamefont {Shyue~Ping}\ \bibnamefont {Ong}},\ }\bibfield
  {title} {Graph Networks as a Universal Machine Learning Framework for
  Molecules and Crystals,\ }\href {\doibase 10.1021/acs.chemmater.9b01294}
  {\bibfield  {journal} {\bibinfo  {journal} {Chemistry of Materials}\ }\textbf
  {\bibinfo {volume} {31}},\ \bibinfo {pages} {3564--3572} (\bibinfo {year}
  {2019})},\ \Eprint
  {http://arxiv.org/abs/https://doi.org/10.1021/acs.chemmater.9b01294}
  {https://doi.org/10.1021/acs.chemmater.9b01294} \BibitemShut {NoStop}%
\bibitem [{\citenamefont {Ward}\ \emph {et~al.}(2018)\citenamefont {Ward},
  \citenamefont {Dunn}, \citenamefont {Faghaninia}, \citenamefont {Zimmermann},
  \citenamefont {Bajaj}, \citenamefont {Wang}, \citenamefont {Montoya},
  \citenamefont {Chen}, \citenamefont {Bystrom}, \citenamefont {Dylla},
  \citenamefont {Chard}, \citenamefont {Asta}, \citenamefont {Persson},
  \citenamefont {Snyder}, \citenamefont {Foster},\ and\ \citenamefont
  {Jain}}]{matminer}%
  \BibitemOpen
  \bibfield  {author} {\bibinfo {author} {\bibfnamefont {Logan}\ \bibnamefont
  {Ward}}, \bibinfo {author} {\bibfnamefont {Alexander}\ \bibnamefont {Dunn}},
  \bibinfo {author} {\bibfnamefont {Alireza}\ \bibnamefont {Faghaninia}},
  \bibinfo {author} {\bibfnamefont {Nils~E.R.}\ \bibnamefont {Zimmermann}},
  \bibinfo {author} {\bibfnamefont {Saurabh}\ \bibnamefont {Bajaj}}, \bibinfo
  {author} {\bibfnamefont {Qi}~\bibnamefont {Wang}}, \bibinfo {author}
  {\bibfnamefont {Joseph}\ \bibnamefont {Montoya}}, \bibinfo {author}
  {\bibfnamefont {Jiming}\ \bibnamefont {Chen}}, \bibinfo {author}
  {\bibfnamefont {Kyle}\ \bibnamefont {Bystrom}}, \bibinfo {author}
  {\bibfnamefont {Maxwell}\ \bibnamefont {Dylla}}, \bibinfo {author}
  {\bibfnamefont {Kyle}\ \bibnamefont {Chard}}, \bibinfo {author}
  {\bibfnamefont {Mark}\ \bibnamefont {Asta}}, \bibinfo {author} {\bibfnamefont
  {Kristin~A.}\ \bibnamefont {Persson}}, \bibinfo {author} {\bibfnamefont
  {G.~Jeffrey}\ \bibnamefont {Snyder}}, \bibinfo {author} {\bibfnamefont {Ian}\
  \bibnamefont {Foster}}, \ and\ \bibinfo {author} {\bibfnamefont {Anubhav}\
  \bibnamefont {Jain}},\ }\bibfield  {title} {Matminer: An open source toolkit
  for materials data mining,\ }\href {\doibase
  https://doi.org/10.1016/j.commatsci.2018.05.018} {\bibfield  {journal}
  {\bibinfo  {journal} {Computational Materials Science}\ }\textbf {\bibinfo
  {volume} {152}},\ \bibinfo {pages} {60--69} (\bibinfo {year}
  {2018})}\BibitemShut {NoStop}%
\bibitem [{\citenamefont {Ward}\ \emph {et~al.}(2016)\citenamefont {Ward},
  \citenamefont {Agrawal}, \citenamefont {Choudhary},\ and\ \citenamefont
  {Wolverton}}]{Ward:2016aa}%
  \BibitemOpen
  \bibfield  {author} {\bibinfo {author} {\bibfnamefont {Logan}\ \bibnamefont
  {Ward}}, \bibinfo {author} {\bibfnamefont {Ankit}\ \bibnamefont {Agrawal}},
  \bibinfo {author} {\bibfnamefont {Alok}\ \bibnamefont {Choudhary}}, \ and\
  \bibinfo {author} {\bibfnamefont {Christopher}\ \bibnamefont {Wolverton}},\
  }\bibfield  {title} {A general-purpose machine learning framework for
  predicting properties of inorganic materials,\ }\href {\doibase
  10.1038/npjcompumats.2016.28} {\bibfield  {journal} {\bibinfo  {journal} {npj
  Computational Materials}\ }\textbf {\bibinfo {volume} {2}},\ \bibinfo {pages}
  {16028} (\bibinfo {year} {2016})}\BibitemShut {NoStop}%
\bibitem [{\citenamefont {Chakoumakos}\ \emph {et~al.}(2000)\citenamefont
  {Chakoumakos}, \citenamefont {Sales}, \citenamefont {Mandrus},\ and\
  \citenamefont {Nolas}}]{CHAKOUMAKOS200080}%
  \BibitemOpen
  \bibfield  {author} {\bibinfo {author} {\bibfnamefont {B.C.}\ \bibnamefont
  {Chakoumakos}}, \bibinfo {author} {\bibfnamefont {B.C.}\ \bibnamefont
  {Sales}}, \bibinfo {author} {\bibfnamefont {D.G.}\ \bibnamefont {Mandrus}}, \
  and\ \bibinfo {author} {\bibfnamefont {G.S.}\ \bibnamefont {Nolas}},\
  }\bibfield  {title} {Structural disorder and thermal conductivity of the
  semiconducting clathrate {Sr8Ga16Ge30},\ }\href {\doibase
  https://doi.org/10.1016/S0925-8388(99)00531-9} {\bibfield  {journal}
  {\bibinfo  {journal} {Journal of Alloys and Compounds}\ }\textbf {\bibinfo
  {volume} {296}},\ \bibinfo {pages} {80--86} (\bibinfo {year}
  {2000})}\BibitemShut {NoStop}%
\bibitem [{\citenamefont {Larkin}\ and\ \citenamefont
  {McGaughey}(2014)}]{PhysRevB.89.144303}%
  \BibitemOpen
  \bibfield  {author} {\bibinfo {author} {\bibfnamefont {Jason~M.}\
  \bibnamefont {Larkin}}\ and\ \bibinfo {author} {\bibfnamefont {Alan J.~H.}\
  \bibnamefont {McGaughey}},\ }\bibfield  {title} {Thermal conductivity
  accumulation in amorphous silica and amorphous silicon,\ }\href {\doibase
  10.1103/PhysRevB.89.144303} {\bibfield  {journal} {\bibinfo  {journal} {Phys.
  Rev. B}\ }\textbf {\bibinfo {volume} {89}},\ \bibinfo {pages} {144303}
  (\bibinfo {year} {2014})}\BibitemShut {NoStop}%
\bibitem [{\citenamefont {Chiritescu}\ \emph {et~al.}(2007)\citenamefont
  {Chiritescu}, \citenamefont {Cahill}, \citenamefont {Nguyen}, \citenamefont
  {Johnson}, \citenamefont {Bodapati}, \citenamefont {Keblinski},\ and\
  \citenamefont {Zschack}}]{chiritescu2007ultralow}%
  \BibitemOpen
  \bibfield  {author} {\bibinfo {author} {\bibfnamefont {Catalin}\ \bibnamefont
  {Chiritescu}}, \bibinfo {author} {\bibfnamefont {David~G}\ \bibnamefont
  {Cahill}}, \bibinfo {author} {\bibfnamefont {Ngoc}\ \bibnamefont {Nguyen}},
  \bibinfo {author} {\bibfnamefont {David}\ \bibnamefont {Johnson}}, \bibinfo
  {author} {\bibfnamefont {Arun}\ \bibnamefont {Bodapati}}, \bibinfo {author}
  {\bibfnamefont {Pawel}\ \bibnamefont {Keblinski}}, \ and\ \bibinfo {author}
  {\bibfnamefont {Paul}\ \bibnamefont {Zschack}},\ }\bibfield  {title}
  {Ultralow thermal conductivity in disordered, layered {WSe2} crystals,\
  }\href@noop {} {\bibfield  {journal} {\bibinfo  {journal} {Science}\ }\textbf
  {\bibinfo {volume} {315}},\ \bibinfo {pages} {351--353} (\bibinfo {year}
  {2007})}\BibitemShut {NoStop}%
\bibitem [{\citenamefont {Xia}\ \emph {et~al.}(2020{\natexlab{b}})\citenamefont
  {Xia}, \citenamefont {Pal}, \citenamefont {He}, \citenamefont {Ozoli\ifmmode
  \mbox{\c{n}}\else \c{n}\fi{}\ifmmode~\check{s}\else \v{s}\fi{}},\ and\
  \citenamefont {Wolverton}}]{tvs2020}%
  \BibitemOpen
  \bibfield  {author} {\bibinfo {author} {\bibfnamefont {Yi}~\bibnamefont
  {Xia}}, \bibinfo {author} {\bibfnamefont {Koushik}\ \bibnamefont {Pal}},
  \bibinfo {author} {\bibfnamefont {Jiangang}\ \bibnamefont {He}}, \bibinfo
  {author} {\bibfnamefont {Vidvuds}\ \bibnamefont {Ozoli\ifmmode
  \mbox{\c{n}}\else \c{n}\fi{}\ifmmode~\check{s}\else \v{s}\fi{}}}, \ and\
  \bibinfo {author} {\bibfnamefont {Chris}\ \bibnamefont {Wolverton}},\
  }\bibfield  {title} {Particlelike Phonon Propagation Dominates Ultralow
  Lattice Thermal Conductivity in Crystalline {Tl}$_{3}${VSe}$_{4}$,\ }\href
  {\doibase 10.1103/PhysRevLett.124.065901} {\bibfield  {journal} {\bibinfo
  {journal} {Phys. Rev. Lett.}\ }\textbf {\bibinfo {volume} {124}},\ \bibinfo
  {pages} {065901} (\bibinfo {year} {2020}{\natexlab{b}})}\BibitemShut
  {NoStop}%
\bibitem [{\citenamefont {Xia}\ \emph {et~al.}(2020{\natexlab{c}})\citenamefont
  {Xia}, \citenamefont {Hegde}, \citenamefont {Pal}, \citenamefont {Hua},
  \citenamefont {Gaines}, \citenamefont {Patel}, \citenamefont {He},
  \citenamefont {Aykol},\ and\ \citenamefont {Wolverton}}]{rczb2020}%
  \BibitemOpen
  \bibfield  {author} {\bibinfo {author} {\bibfnamefont {Yi}~\bibnamefont
  {Xia}}, \bibinfo {author} {\bibfnamefont {Vinay~I.}\ \bibnamefont {Hegde}},
  \bibinfo {author} {\bibfnamefont {Koushik}\ \bibnamefont {Pal}}, \bibinfo
  {author} {\bibfnamefont {Xia}\ \bibnamefont {Hua}}, \bibinfo {author}
  {\bibfnamefont {Dale}\ \bibnamefont {Gaines}}, \bibinfo {author}
  {\bibfnamefont {Shane}\ \bibnamefont {Patel}}, \bibinfo {author}
  {\bibfnamefont {Jiangang}\ \bibnamefont {He}}, \bibinfo {author}
  {\bibfnamefont {Muratahan}\ \bibnamefont {Aykol}}, \ and\ \bibinfo {author}
  {\bibfnamefont {Chris}\ \bibnamefont {Wolverton}},\ }\bibfield  {title}
  {High-Throughput Study of Lattice Thermal Conductivity in Binary Rocksalt and
  Zinc Blende Compounds Including Higher-Order Anharmonicity,\ }\href {\doibase
  10.1103/PhysRevX.10.041029} {\bibfield  {journal} {\bibinfo  {journal} {Phys.
  Rev. X}\ }\textbf {\bibinfo {volume} {10}},\ \bibinfo {pages} {041029}
  (\bibinfo {year} {2020}{\natexlab{c}})}\BibitemShut {NoStop}%
\bibitem [{\citenamefont {Sun}\ and\ \citenamefont {Allen}(2010)}]{Sun2010}%
  \BibitemOpen
  \bibfield  {author} {\bibinfo {author} {\bibfnamefont {Tao}\ \bibnamefont
  {Sun}}\ and\ \bibinfo {author} {\bibfnamefont {Philip~B.}\ \bibnamefont
  {Allen}},\ }\bibfield  {title} {Lattice thermal conductivity: Computations
  and theory of the high-temperature breakdown of the phonon-gas model,\ }\href
  {\doibase 10.1103/PhysRevB.82.224305} {\bibfield  {journal} {\bibinfo
  {journal} {Phys. Rev. B}\ }\textbf {\bibinfo {volume} {82}},\ \bibinfo
  {pages} {224305} (\bibinfo {year} {2010})}\BibitemShut {NoStop}%
\bibitem [{\citenamefont {Delaire}\ \emph {et~al.}(2011)\citenamefont
  {Delaire}, \citenamefont {Ma}, \citenamefont {Marty}, \citenamefont {May},
  \citenamefont {McGuire}, \citenamefont {Du}, \citenamefont {Singh},
  \citenamefont {Podlesnyak}, \citenamefont {Ehlers}, \citenamefont {Lumsden}
  \emph {et~al.}}]{delaire2011giant}%
  \BibitemOpen
  \bibfield  {author} {\bibinfo {author} {\bibfnamefont {Olivier}\ \bibnamefont
  {Delaire}}, \bibinfo {author} {\bibfnamefont {Jie}\ \bibnamefont {Ma}},
  \bibinfo {author} {\bibfnamefont {Karol}\ \bibnamefont {Marty}}, \bibinfo
  {author} {\bibfnamefont {Andrew~F}\ \bibnamefont {May}}, \bibinfo {author}
  {\bibfnamefont {Michael~A}\ \bibnamefont {McGuire}}, \bibinfo {author}
  {\bibfnamefont {Mao-Hua}\ \bibnamefont {Du}}, \bibinfo {author}
  {\bibfnamefont {David~J}\ \bibnamefont {Singh}}, \bibinfo {author}
  {\bibfnamefont {A}~\bibnamefont {Podlesnyak}}, \bibinfo {author}
  {\bibfnamefont {G}~\bibnamefont {Ehlers}}, \bibinfo {author} {\bibfnamefont
  {MD}~\bibnamefont {Lumsden}},  \emph {et~al.},\ }\bibfield  {title} {Giant
  anharmonic phonon scattering in {PbTe},\ }\href@noop {} {\bibfield  {journal}
  {\bibinfo  {journal} {Nature materials}\ }\textbf {\bibinfo {volume} {10}},\
  \bibinfo {pages} {614--619} (\bibinfo {year} {2011})}\BibitemShut {NoStop}%
\bibitem [{\citenamefont {Kresse}\ and\ \citenamefont {Hafner}(1993)}]{Vasp1}%
  \BibitemOpen
  \bibfield  {author} {\bibinfo {author} {\bibfnamefont {G.}~\bibnamefont
  {Kresse}}\ and\ \bibinfo {author} {\bibfnamefont {J.}~\bibnamefont
  {Hafner}},\ }\bibfield  {title} {\textit{Ab initio} molecular dynamics for
  liquid metals,\ }\href {\doibase 10.1103/PhysRevB.47.558} {\bibfield
  {journal} {\bibinfo  {journal} {Phys. Rev. B}\ }\textbf {\bibinfo {volume}
  {47}},\ \bibinfo {pages} {558--561} (\bibinfo {year} {1993})}\BibitemShut
  {NoStop}%
\bibitem [{\citenamefont {Kresse}\ and\ \citenamefont {Hafner}(1994)}]{Vasp2}%
  \BibitemOpen
  \bibfield  {author} {\bibinfo {author} {\bibfnamefont {G.}~\bibnamefont
  {Kresse}}\ and\ \bibinfo {author} {\bibfnamefont {J.}~\bibnamefont
  {Hafner}},\ }\bibfield  {title} {\textit{Ab initio} molecular-dynamics
  simulation of the liquid-metal\char21{}amorphous-semiconductor transition in
  germanium,\ }\href {\doibase 10.1103/PhysRevB.49.14251} {\bibfield  {journal}
  {\bibinfo  {journal} {Phys. Rev. B}\ }\textbf {\bibinfo {volume} {49}},\
  \bibinfo {pages} {14251--14269} (\bibinfo {year} {1994})}\BibitemShut
  {NoStop}%
\bibitem [{\citenamefont {Kresse}\ and\ \citenamefont
  {Furthm{\"u}ller}(1996)}]{Vasp3}%
  \BibitemOpen
  \bibfield  {author} {\bibinfo {author} {\bibfnamefont {G.}~\bibnamefont
  {Kresse}}\ and\ \bibinfo {author} {\bibfnamefont {J.}~\bibnamefont
  {Furthm{\"u}ller}},\ }\bibfield  {title} {Efficiency of ab-initio total
  energy calculations for metals and semiconductors using a plane-wave basis
  set,\ }\href@noop {} {\bibfield  {journal} {\bibinfo  {journal} {Comput.
  Mater. Sci.}\ }\textbf {\bibinfo {volume} {6}},\ \bibinfo {pages} {15--50}
  (\bibinfo {year} {1996})}\BibitemShut {NoStop}%
\bibitem [{\citenamefont {Kresse}\ and\ \citenamefont
  {Furthm\"uller}(1996)}]{Vasp4}%
  \BibitemOpen
  \bibfield  {author} {\bibinfo {author} {\bibfnamefont {G.}~\bibnamefont
  {Kresse}}\ and\ \bibinfo {author} {\bibfnamefont {J.}~\bibnamefont
  {Furthm\"uller}},\ }\bibfield  {title} {Efficient iterative schemes for
  \textit{ab initio} total-energy calculations using a plane-wave basis set,\
  }\href {\doibase 10.1103/PhysRevB.54.11169} {\bibfield  {journal} {\bibinfo
  {journal} {Phys. Rev. B}\ }\textbf {\bibinfo {volume} {54}},\ \bibinfo
  {pages} {11169--11186} (\bibinfo {year} {1996})}\BibitemShut {NoStop}%
\bibitem [{\citenamefont {Bl\"ochl}(1994)}]{paw}%
  \BibitemOpen
  \bibfield  {author} {\bibinfo {author} {\bibfnamefont {P.~E.}\ \bibnamefont
  {Bl\"ochl}},\ }\bibfield  {title} {Projector augmented-wave method,\ }\href
  {\doibase 10.1103/PhysRevB.50.17953} {\bibfield  {journal} {\bibinfo
  {journal} {Phys. Rev. B}\ }\textbf {\bibinfo {volume} {50}},\ \bibinfo
  {pages} {17953--17979} (\bibinfo {year} {1994})}\BibitemShut {NoStop}%
\bibitem [{\citenamefont {Perdew}\ \emph {et~al.}(2008)\citenamefont {Perdew},
  \citenamefont {Ruzsinszky}, \citenamefont {Csonka}, \citenamefont {Vydrov},
  \citenamefont {Scuseria}, \citenamefont {Constantin}, \citenamefont {Zhou},\
  and\ \citenamefont {Burke}}]{Perdew2008}%
  \BibitemOpen
  \bibfield  {author} {\bibinfo {author} {\bibfnamefont {John~P.}\ \bibnamefont
  {Perdew}}, \bibinfo {author} {\bibfnamefont {Adrienn}\ \bibnamefont
  {Ruzsinszky}}, \bibinfo {author} {\bibfnamefont {G\'abor~I.}\ \bibnamefont
  {Csonka}}, \bibinfo {author} {\bibfnamefont {Oleg~A.}\ \bibnamefont
  {Vydrov}}, \bibinfo {author} {\bibfnamefont {Gustavo~E.}\ \bibnamefont
  {Scuseria}}, \bibinfo {author} {\bibfnamefont {Lucian~A.}\ \bibnamefont
  {Constantin}}, \bibinfo {author} {\bibfnamefont {Xiaolan}\ \bibnamefont
  {Zhou}}, \ and\ \bibinfo {author} {\bibfnamefont {Kieron}\ \bibnamefont
  {Burke}},\ }\bibfield  {title} {Restoring the Density-Gradient Expansion for
  Exchange in Solids and Surfaces,\ }\href {\doibase
  10.1103/PhysRevLett.100.136406} {\bibfield  {journal} {\bibinfo  {journal}
  {Phys. Rev. Lett.}\ }\textbf {\bibinfo {volume} {100}},\ \bibinfo {pages}
  {136406} (\bibinfo {year} {2008})}\BibitemShut {NoStop}%
\bibitem [{\citenamefont {Perdew}\ \emph {et~al.}(1996)\citenamefont {Perdew},
  \citenamefont {Burke},\ and\ \citenamefont {Wang}}]{gga}%
  \BibitemOpen
  \bibfield  {author} {\bibinfo {author} {\bibfnamefont {John~P.}\ \bibnamefont
  {Perdew}}, \bibinfo {author} {\bibfnamefont {Kieron}\ \bibnamefont {Burke}},
  \ and\ \bibinfo {author} {\bibfnamefont {Yue}\ \bibnamefont {Wang}},\
  }\bibfield  {title} {Generalized gradient approximation for the
  exchange-correlation hole of a many-electron system,\ }\href {\doibase
  10.1103/PhysRevB.54.16533} {\bibfield  {journal} {\bibinfo  {journal} {Phys.
  Rev. B}\ }\textbf {\bibinfo {volume} {54}},\ \bibinfo {pages} {16533--16539}
  (\bibinfo {year} {1996})}\BibitemShut {NoStop}%
\bibitem [{\citenamefont {Hohenberg}\ and\ \citenamefont {Kohn}(1964)}]{dft}%
  \BibitemOpen
  \bibfield  {author} {\bibinfo {author} {\bibfnamefont {P.}~\bibnamefont
  {Hohenberg}}\ and\ \bibinfo {author} {\bibfnamefont {W.}~\bibnamefont
  {Kohn}},\ }\bibfield  {title} {Inhomogeneous Electron Gas,\ }\href {\doibase
  10.1103/PhysRev.136.B864} {\bibfield  {journal} {\bibinfo  {journal} {Phys.
  Rev.}\ }\textbf {\bibinfo {volume} {136}},\ \bibinfo {pages} {B864--B871}
  (\bibinfo {year} {1964})}\BibitemShut {NoStop}%
\bibitem [{\citenamefont {Li}\ \emph {et~al.}(2014)\citenamefont {Li},
  \citenamefont {Carrete}, \citenamefont {Katcho},\ and\ \citenamefont
  {Mingo}}]{shengbte}%
  \BibitemOpen
  \bibfield  {author} {\bibinfo {author} {\bibfnamefont {Wu}~\bibnamefont
  {Li}}, \bibinfo {author} {\bibfnamefont {Jes{\'u}s}\ \bibnamefont {Carrete}},
  \bibinfo {author} {\bibfnamefont {Nebil~A.}\ \bibnamefont {Katcho}}, \ and\
  \bibinfo {author} {\bibfnamefont {Natalio}\ \bibnamefont {Mingo}},\
  }\bibfield  {title} {ShengBTE: A solver of the Boltzmann transport equation
  for phonons,\ }\href {\doibase http://dx.doi.org/10.1016/j.cpc.2014.02.015}
  {\bibfield  {journal} {\bibinfo  {journal} {Computer Physics Communications}\
  }\textbf {\bibinfo {volume} {185}},\ \bibinfo {pages} {1747 -- 1758}
  (\bibinfo {year} {2014})}\BibitemShut {NoStop}%
\bibitem [{\citenamefont {Errea}\ \emph {et~al.}(2014)\citenamefont {Errea},
  \citenamefont {Calandra},\ and\ \citenamefont {Mauri}}]{Errea2014}%
  \BibitemOpen
  \bibfield  {author} {\bibinfo {author} {\bibfnamefont {Ion}\ \bibnamefont
  {Errea}}, \bibinfo {author} {\bibfnamefont {Matteo}\ \bibnamefont
  {Calandra}}, \ and\ \bibinfo {author} {\bibfnamefont {Francesco}\
  \bibnamefont {Mauri}},\ }\bibfield  {title} {Anharmonic free energies and
  phonon dispersions from the stochastic self-consistent harmonic
  approximation: Application to platinum and palladium hydrides,\ }\href
  {\doibase 10.1103/PhysRevB.89.064302} {\bibfield  {journal} {\bibinfo
  {journal} {Phys. Rev. B}\ }\textbf {\bibinfo {volume} {89}},\ \bibinfo
  {pages} {064302} (\bibinfo {year} {2014})}\BibitemShut {NoStop}%
\bibitem [{\citenamefont {Tadano}\ and\ \citenamefont
  {Tsuneyuki}(2015)}]{Tadano2015}%
  \BibitemOpen
  \bibfield  {author} {\bibinfo {author} {\bibfnamefont {Terumasa}\
  \bibnamefont {Tadano}}\ and\ \bibinfo {author} {\bibfnamefont {Shinji}\
  \bibnamefont {Tsuneyuki}},\ }\bibfield  {title} {Self-consistent phonon
  calculations of lattice dynamical properties in cubic {SrTiO}$_{3}$ with
  first-principles anharmonic force constants,\ }\href {\doibase
  10.1103/PhysRevB.92.054301} {\bibfield  {journal} {\bibinfo  {journal} {Phys.
  Rev. B}\ }\textbf {\bibinfo {volume} {92}},\ \bibinfo {pages} {054301}
  (\bibinfo {year} {2015})}\BibitemShut {NoStop}%
\bibitem [{\citenamefont {Saal}\ \emph {et~al.}(2013)\citenamefont {Saal},
  \citenamefont {Kirklin}, \citenamefont {Aykol}, \citenamefont {Meredig},\
  and\ \citenamefont {Wolverton}}]{Saal:2013aa}%
  \BibitemOpen
  \bibfield  {author} {\bibinfo {author} {\bibfnamefont {James~E.}\
  \bibnamefont {Saal}}, \bibinfo {author} {\bibfnamefont {Scott}\ \bibnamefont
  {Kirklin}}, \bibinfo {author} {\bibfnamefont {Muratahan}\ \bibnamefont
  {Aykol}}, \bibinfo {author} {\bibfnamefont {Bryce}\ \bibnamefont {Meredig}},
  \ and\ \bibinfo {author} {\bibfnamefont {C.}~\bibnamefont {Wolverton}},\
  }\bibfield  {title} {{Materials Design and Discovery with High-Throughput
  Density Functional Theory: The Open Quantum Materials Database (OQMD)},\
  }\href {\doibase 10.1007/s11837-013-0755-4} {\bibfield  {journal} {\bibinfo
  {journal} {JOM}\ }\textbf {\bibinfo {volume} {65}},\ \bibinfo {pages}
  {1501--1509} (\bibinfo {year} {2013})}\BibitemShut {NoStop}%
\bibitem [{\citenamefont {Kirklin}\ \emph {et~al.}(2015)\citenamefont
  {Kirklin}, \citenamefont {Saal}, \citenamefont {Meredig}, \citenamefont
  {Thompson}, \citenamefont {Doak}, \citenamefont {Aykol}, \citenamefont
  {R{\"u}hl},\ and\ \citenamefont {Wolverton}}]{Kirklin:2015aa}%
  \BibitemOpen
  \bibfield  {author} {\bibinfo {author} {\bibfnamefont {Scott}\ \bibnamefont
  {Kirklin}}, \bibinfo {author} {\bibfnamefont {James~E}\ \bibnamefont {Saal}},
  \bibinfo {author} {\bibfnamefont {Bryce}\ \bibnamefont {Meredig}}, \bibinfo
  {author} {\bibfnamefont {Alex}\ \bibnamefont {Thompson}}, \bibinfo {author}
  {\bibfnamefont {Jeff~W}\ \bibnamefont {Doak}}, \bibinfo {author}
  {\bibfnamefont {Muratahan}\ \bibnamefont {Aykol}}, \bibinfo {author}
  {\bibfnamefont {Stephan}\ \bibnamefont {R{\"u}hl}}, \ and\ \bibinfo {author}
  {\bibfnamefont {Chris}\ \bibnamefont {Wolverton}},\ }\bibfield  {title} {{The
  Open Quantum Materials Database (OQMD)}: assessing the accuracy of {DFT}
  formation energies,\ }\href {\doibase 10.1038/npjcompumats.2015.10}
  {\bibfield  {journal} {\bibinfo  {journal} {npj Computational Materials}\
  }\textbf {\bibinfo {volume} {1}},\ \bibinfo {pages} {15010} (\bibinfo {year}
  {2015})}\BibitemShut {NoStop}%
\end{thebibliography}%

\end{document}